\documentclass[aps,prd,onecolumn,showpacs,groupedaddress,nofootinbib]{revtex4}  
\usepackage{graphicx}
\usepackage{epstopdf}
\usepackage{amsmath}
\usepackage{amsfonts}
\usepackage{amssymb}
\usepackage{appendix}
\usepackage{comment}
\usepackage{bbold}
\usepackage{color}
\usepackage{slashed}
\usepackage{subfigure}
\usepackage{wrapfig}
\usepackage{setspace}
\usepackage{footnote}
\usepackage{multirow}
\usepackage{braket}
\usepackage[normalem]{ulem}
\usepackage[a4paper, portrait, margin=0.5in]{geometry}
\usepackage[utf8]{inputenc}
\usepackage{mathrsfs}
\usepackage{bbm}
\allowdisplaybreaks

\usepackage{url}
\usepackage{wrapfig}
\usepackage{braket}

\begin{document}
\singlespacing
{\hfill NUHEP-TH/17-01}

\title{Searches for new physics at the Hyper-Kamiokande experiment}

\author{Kevin J. Kelly}
\affiliation{Northwestern University, Department of Physics \& Astronomy, 2145 Sheridan Road, Evanston, IL 60208, USA}

\begin{abstract}
We investigate the ability of the upcoming Hyper-Kamiokande (Hyper-K) neutrino experiment to detect new physics phenomena beyond the standard, three-massive-neutrinos paradigm; namely the existence of a fourth, sterile neutrino or weaker-than-weak, non-standard neutrino interactions. With both beam-based neutrinos from the Japan Proton Accelerator Research Complex (J-PARC) and atmospheric neutrinos, Hyper-K is capable of exploring new ranges of parameter space in these new-physics scenarios. We find that Hyper-K has comparable capability to the upcoming Deep Underground Neutrino Experiment (DUNE), and that combining both beam- and atmospheric-based data can clear up degeneracies in the parameter spaces of interest. We also comment on the potential improvement in searches for new physics if a combined analysis were performed using Hyper-K and DUNE data.
\end{abstract}

\maketitle

\section{Introduction}
\label{sec:Introduction}

With the discovery that neutrinos have mass and leptons mix, neutrino oscillations have been identified as a clear direction to study physics beyond the Standard Model (SM). Many existing experiments have measured the neutrino mass splittings and the leptonic mixing matrix, and several next-generation experiments, such as the Hyper-Kamiokande Experiment (Hyper-K)~\cite{Abe:2015zbg,Abe:2016ero} and the Deep Underground Neutrino Experiment (DUNE)~\cite{Adams:2013qkq,Acciarri:2015uup}, have been proposed to continue this study at long baselines. Hyper-K and DUNE aim to answer several remaining questions regarding lepton mixing with three SM neutrinos that have mass (which we will refer to as the ``three-massive-neutrinos paradigm''). In addition to this, the next generation experiments will be able to test for physics beyond the three-massive-neutrinos paradigm.

Many hypotheses exist that extend beyond the three-massive-neutrinos paradigm that are still consistent with present data. Among these are the proposal that the leptonic mixing matrix is non-unitary (unlike the quark mixing matrix, the unitarity of the leptonic mixing matrix is not well-constrained~\cite{Antusch:2006vwa,Qian:2013ora,Parke:2015goa}), the existence of singlet fermion fields propagating in large extra dimensions, the addition of a fourth neutrino state, and the existence of interactions involving neutrinos aside from the weak interactions. In this work, we will focus on the last two, referred to respectively as the sterile neutrino and non-standard neutrino interaction hypotheses.

The addition of a fourth, sterile neutrino as an extension to the three-massive-neutrinos paradigm has been studied extensively in the literature -- theoretical motivations for a fourth neutrino are wide-ranging, from explaining the mechanism by which the light neutrinos acquire a mass (see, e.g., Ref.~\cite{Caldwell:1993kn}), to alleviating experimental oscillation results that appear inconsistent with the three-massive-neutrinos paradigm~\cite{Aguilar:2001ty,AguilarArevalo:2008rc,Mention:2011rk,Frekers:2011zz,Aguilar-Arevalo:2012fmn,Aguilar-Arevalo:2013pmq}. These motivations require a fourth neutrino with widely varying mass -- in this work we will focus on cases with a new mass eigenstate $m_4 \lesssim 1$ eV which can impact neutrino oscillations at long baselines. Sterile neutrinos in this mass range have been studied in the context of short-baseline oscillations in Refs.~\cite{deGouvea:2014aoa,Adey:2015iha,Gariazzo:2015rra,Giunti:2015wnd,Choubey:2016fpi}, and Refs.~\cite{Donini:2007yf,Dighe:2007uf,deGouvea:2008qk,Meloni:2010zr,Bhattacharya:2011ee,Hollander:2014iha,Berryman:2015nua,Tabrizi:2015bba,Gariazzo:2015rra,Gandhi:2015xza,Palazzo:2015gja,deGouvea:2015euy,Giunti:2015wnd,Agarwalla:2016mrc,Agarwalla:2016xxa,Dutta:2016glq,Blennow:2016jkn} have studied the impact of a sterile neutrino in long-baseline oscillations, as this work will. Constraints on a fourth neutrino over a wider range of masses have been discussed in Refs.~\cite{Atre:2009rg,Vincent:2014rja,Drewes:2015iva,Deppisch:2015qwa,deGouvea:2015euy,Adhikari:2016bei}. 

Non-standard neutrino interactions (NSI), originally proposed as a solution to the solar neutrino problem~\cite{Wolfenstein:1977ue}, have been studied in a number of situations, all of which introduce additional interactions involving neutrinos and other fermions. Refs.~\cite{Guzzo:1991hi,Krastev:1992zx,Friedland:2004pp,Miranda:2004nb,Bolanos:2008km,Palazzo:2009rb,Escrihuela:2009up} have studied the impact of NSI on solar neutrino oscillations, Refs.~\cite{GonzalezGarcia:1998hj,Fornengo:1999zp,Fornengo:2001pm,Huber:2001zw,Friedland:2004ah,Friedland:2005vy,Yasuda:2010hw,GonzalezGarcia:2011my,Esmaili:2013fva,Choubey:2014iia,Mocioiu:2014gua,Fukasawa:2015jaa,Choubey:2015xha,Salvado:2016uqu} have studied how they contribute to atmospheric neutrino oscillations, and Refs.~\cite{Friedland:2006pi,Blennow:2007pu,EstebanPretel:2008qi,Kopp:2010qt,Coloma:2011rq,Friedland:2012tq,Coelho:2012bp,Adamson:2013ovz,Girardi:2014kca,Blennow:2015nxa,Masud:2015xva,deGouvea:2015ndi,Coloma:2015kiu,Liao:2016hsa,Forero:2016cmb,Huitu:2016bmb,Bakhti:2016prn,Masud:2016bvp,Miranda:2016wdr,Coloma:2016gei,Khan:2016uon,deGouvea:2016pom,Masud:2016gcl,Blennow:2016etl,Bakhti:2016gic,Farzan:2016wym,Forero:2016ghr,Fukasawa:2016gvm,Blennow:2016jkn,Deepthi:2016erc} have studied NSI in the context of accelerator-based neutrino oscillations, particularly focusing on the upcoming long-baseline oscillation experiments. Recently, NSI have been discussed regarding the Hyper-Kamiokande experiment in Refs.~\cite{Coloma:2015kiu,Ge:2016dlx,Fukasawa:2016lew,Fukasawa:2016nwn,Liao:2016orc,Rout:2017udo,Ghosh:2017ged}. This work adds to the discussion of NSI at Hyper-K by conducting a thorough, multi-parameter analysis of the sensitivity of the experiment, utilizing both its beam- and atmospheric-based capabilities.

This manuscript is organized as follows: in Section~\ref{sec:Osc}, we introduce the oscillation formalism used when discussing the three-massive-neutrinos paradigm, as well as the extensions to this: sterile neutrinos and non-standard neutrino interactions. In Section~\ref{sec:HK}, we discuss the capabilities of the Hyper-Kamiokande experiment, in both the detection of neutrinos generated from the Japan Proton Accelerator Research Complex (J-PARC) and the detection of neutrinos produced in the atmosphere. Here, we also discuss our analysis method. In Section~\ref{sec:Results}, we present the results of our analyses, including the ability of the Hyper-Kamiokande experiment to detect sterile neutrinos and non-standard interactions, and in Section~\ref{sec:Conclusions}, we offer some concluding remarks.

\section{Oscillations and new Neutrino Physics}
\label{sec:Osc}
We direct the reader to, for example, Refs.~\cite{Berryman:2015nua} and \cite{deGouvea:2015ndi} for more thorough discussions on long-baseline neutrino oscillations regarding four-neutrino scenarios and non-standard neutrino interactions (NSI) respectively. Here, we explain the three-massive-neutrinos paradigm and three-neutrino oscillations, and in Secs.~\ref{subsec:Sterile} and \ref{subsec:NSI}, we introduce the formalisms regarding oscillations with four neutrinos and with NSI, respectively.

With three neutrinos and two non-zero mass-squared splittings $\Delta m_{ij}^2 \equiv m_j^2 - m_i^2$, we characterize oscillations using a $3\times 3$ unitary, PMNS matrix $U$. This requires three mixing angles ($\theta_{12}$, $\theta_{13}$, and $\theta_{23}$) and one $CP$-violating phase ($\delta$) to describe oscillations. We use the Particle Data Group convention for $U$~\cite{Olive:2016xmw}. The mixing angles and mass splittings have been measured to be non-zero, but two important measurements remain: the value of $\delta$ and the mass hierarchy, whether $\Delta m_{13}^2 > 0$ (normal hierarchy) or $\Delta m_{13}^2 < 0$ (inverted hierarchy).

With neutrino states in the flavor basis ($e$, $\mu$, $\tau$), the probability for a neutrino of flavor $\alpha$ to propagate a distance $L$ and be detected as flavor $\beta$ is denoted by the amplitude mod-squared
\begin{equation}
P_{\alpha\beta} \equiv |\mathcal{A}_{\alpha\beta}|^2 = \left\lvert\bra{\nu_\beta} U e^{-i H_{ij} L} U^\dagger \ket{\nu_\alpha}\right\rvert^2,\label{eq:Prob}
\end{equation}
where $U$ is the PMNS matrix and $H_{ij}$ is the Hamiltonian in the basis in which propagation in vacuum is diagonal. This equation is only valid when the Hamiltonian is constant over the entire distance $L$, and while the neutrinos remain a coherent superposition of plane waves. In this basis and in the ultra-relativistic approximation, $H_{ij} \equiv 1/(2E_\nu)$ diag$\left\lbrace 0, \Delta m_{12}^2, \Delta m_{13}^2\right\rbrace$, where $E_\nu$ is the neutrino energy. While propagating through earth, interactions between the neutrinos and the electrons, protons, and neutrons introduce an effective interaction potential $V$. As these interactions are mediated by $W-$ and $Z-$ bosons (the same interactions that govern neutrino production and detection), $V$ is diagonal in the flavor basis. With this effective interaction potential, we must augment the propagation Hamiltonian:
\begin{equation}
H_{ij} \longrightarrow H_{ij} + U_{i\alpha}^\dagger V_{\alpha\beta} U_{\beta j},
\end{equation}
where the PMNS matrix is used to rotate the potential into the mass basis. The interactions with protons and neutrons are identical between $\nu_\alpha$, $\alpha = e$, $\mu$, $\tau$, and can be absorbed as a phase in the Hamiltonian. The remaining term, coming from $t$-channel interaction between a $\nu_e$ and an electron, mediated by a $W$-boson, is $V_{\alpha\beta} = A$ diag $\left\lbrace 1, 0, 0\right\rbrace$, where $A = \sqrt{2} G_F n_e$. $G_F$ is the Fermi constant, and $n_e$ is the number density of electrons along the path of propagation. For antineutrinos oscillating, $U\to U^*$ and $A\to -A$ (to account for the $s$-channel interaction of $\bar{\nu}_e$ with $e^-$ in matter).

Eq.~(\ref{eq:Prob}) is only valid for an interaction potential $V_{\alpha\beta}$ that is constant over the entire baseline length $L$. For propagation through the earth, the path length and matter density depend strongly on the zenith angle $\theta_z$. We simulate the density profile of the earth to be piecewise constant with four distinct regions ranging from $3$ g/cm$^3$ to $13$ g/cm$^3$, closely resembling the PREM earth density model~\cite{Dziewonski:1981xy}. Eq.~(\ref{eq:Prob}) is then modified, becoming
\begin{equation}
P_{\alpha\beta} = |\mathcal{A}_{\alpha\beta}|^2 = \left\lvert\bra{\nu_\beta} U \left( \prod_{n=1}^N e^{-i H_{ij}^{(n)} L_n} \right) U^\dagger \ket{\nu_\alpha}\right\rvert^2,
\label{eq:AtmosProb}
\end{equation}
where $N$ is the number of distinct regions through which a chord along angle $\theta_z$ passes, $H_{ij}^{(n)}$ is the mass-basis Hamiltonian with the matter density of region $n$, and $L_n$ is the length of the chord through this region.

\begin{table}[!htbp]
\begin{center}
\begin{tabular}{|c||c|c|c|c|c|c|c|}
\hline
Parameter & $\sin^2\theta_{12}$ & $\sin^2\theta_{23}$ & $\sin^2\theta_{13}$ & $\delta$ & $\Delta m_{12}^2$ & $\Delta m_{13}^2$ & $|U_{e2}|^2$ \\ \hline
Value & $0.306$ & $0.441$ & $0.02166$ & $-1.728$ & $\left(7.50^{+0.19}_{-0.17}\right) \times 10^{-5}$ eV$^2$ & $2.524\times 10^{-3}$ eV$^2$ & $0.2994 \pm 0.0117$ \\ \hline
\end{tabular}
\end{center}
\caption{Input values assumed for three-neutrino parameters as extracted from the NuFIT collaboration, Ref.~\cite{Esteban:2016qun}. One-sigma ranges are quoted for $\Delta m_{12}^2$ and $|U_{e2}|^2$ (calculated from measurements of $\sin^2\theta_{12}$ and $\sin^2\theta_{13}$), which are used as priors in later sections.}
\label{tab:NuFitTable}
\end{table}
Unless otherwise specified, we will use the results of the most recent NuFIT calculations (Ref.~\cite{Esteban:2016qun}) as physical values for three-neutrino parameters. These values are listed in Table~\ref{tab:NuFitTable}. We assume that there is a normal mass hierarchy, and do not marginalize over the hierarchy in our analysis. This assumption relies on the measurement of the neutrino mass hierarchy before Hyper-K begins collecting data.

\subsection{Sterile Neutrino}
\label{subsec:Sterile}
While the three-massive-neutrinos paradigm is in agreement with nearly all existing oscillation data, several hints exist that might be explained by a fourth neutrino and a mass splitting of $\Delta m_{14}^2 \simeq 1$ eV$^2$~\cite{Aguilar:2001ty,AguilarArevalo:2008rc,Mention:2011rk,Frekers:2011zz,Aguilar-Arevalo:2012fmn,Aguilar-Arevalo:2013pmq}. Mass splittings in this range are best probed by oscillation experiments with baseline lengths and neutrino energies that satisfy $L/E_\nu \simeq 1$ km/GeV. As they are designed to measure $\Delta m_{13}^2$, long-baseline experiments such as Hyper-K and DUNE are sensitive to lower mass splittings ($\Delta m_{14}^2 \simeq 10^{-2}$ eV$^2$). They also provide a complementary probe to the short-baseline experiments' searches for eV$^2$-scale splittings.

In order to accommodate a fourth neutrino, we must extend the $U(3)$ PMNS matrix into a $U(4)$ matrix. In doing so, we require six mixing angles ($\phi_{ij};$ $i < j;$ $i, j = 1$, $2$, $3$, $4$) and three $CP$-violating phases ($\eta_{i};$ $i = 1,$ $2$, $3$)\footnote{We explicitly label the mixing angles $\phi_{ij}$ and phases $\eta_i$ in the four-neutrino scenario to reduce confusion with the three-massive-neutrinos paradigm. In the limit that $\phi_{i4}\to 0$, $\phi_{12,13,23} = \theta_{12,13,23}$ and the phase $\eta_1 = \delta$.}. Assuming unitarity, the relevant matrix elements are
\begin{align}
U_{e2} =&~ s_{12} c_{13} c_{14}, \\
U_{e3} =&~ s_{13} c_{14} e^{-i \eta_1}, \label{eq:Ue3}\\
U_{e4} =&~ s_{14} e^{-i \eta_2}, \\
U_{\mu 2} =&~ c_{24} \left( c_{12} c_{23} - e^{i\eta_1} s_{12} s_{13} s_{23}\right) - e^{i(\eta_2 - \eta_3)} s_{12} c_{13} s_{14} s_{24}, \\
U_{\mu 3} =&~ s_{23} c_{13} c_{24} - e^{i(\eta_2 - \eta_3 - \eta_1)} s_{13} s_{14} s_{24}, \\
U_{\mu 4} =&~ s_{24} c_{14} e^{-i\eta_3}, \\
U_{\tau 2} =&~ c_{34} \left(-c_{12} s_{23} - e^{i\eta_1} s_{12} s_{13} c_{23}\right) - e^{i\eta_2} c_{13} s_{12} c_{24} s_{14} s_{34} -e^{i\eta_3} \left(c_{12} c_{23} - e^{i\eta_1} s_{12} s_{13} s_{23}\right) s_{24} s_{34}, \\
U_{\tau 3} =&~ c_{13} c_{23} c_{34} - e^{i(\eta_2 - \eta_1)} s_{13} c_{24} s_{14} s_{34}  - e^{i\eta_3} s_{23} c_{13} s_{24} s_{34}, \\
U_{\tau 4} =&~ c_{14} c_{24} s_{34},
\end{align}
where $s_{ij} \equiv \sin{\phi_{ij}}$ and $c_{ij} \equiv \cos{\phi_{ij}}$. The remaining matrix elements may be determined by the unitarity of $U$.

As with the PMNS matrix, the propagation Hamiltonian must be extended. The Hamiltonian in vacuum becomes $H_{ij} = 1/(2E_\nu)$ diag$\left\lbrace 0, \Delta m_{12}^2, \Delta m_{13}^2, \Delta m_{14}^2\right\rbrace$, and the interaction potential for a constant-density environment becomes
\begin{equation}
V_{\alpha\beta} \longrightarrow A \left(\begin{array}{cccc} 1 & 0 & 0 & 0 \\ 0 & 0 & 0 & 0 \\ 0 & 0 & 0 & 0 \\ 0 & 0 & 0 & \frac{n_n}{2 n_e} \end{array} \right),
\end{equation}
where $n_n$ is the number density of neutrons, which we assume to be equal to the number density of electrons in earth. This term comes from the phase removed from the potential discussed above, along with the assumption that the additional eigenstate in the flavor basis is sterile and does not interact with the $W-$ or $Z-$bosons.

As an illustrative example of a sterile neutrino hypothesis, we use the parameters shown in Table~\ref{tab:SterileTable} for comparisons in figures. These parameters are chosen to highlight differences in oscillation probabilities and event yields, and are not used in any of the analyses discussed in Section~\ref{sec:Results}.
\begin{table}[!htbp]
\begin{center}
\begin{tabular}{|c||c|c|c|c|c|c|}
\hline
Parameter & $\sin^2(2\phi_{14})$ & $\sin^2\phi_{24}$ & $\sin^2\phi_{34}$ & $\Delta m_{14}^2$ & $\eta_2$ & $\eta_3$ \\ \hline
Value & $5 \times 10^{-2}$ & $2\times 10^{-2}$ & $0$ & $6 \times 10^{-3}$ eV$^2$ & $0$ & $0$ \\ \hline
\end{tabular}
\end{center}
\caption{Input values used for an illustrative sterile neutrino hypothesis for comparisons in figures throughout this work. The unlisted parameters $\sin^2\phi_{ij}$ are equal to the values $\sin^2\theta_{ij}$ in Table~\ref{tab:NuFitTable} for $i, j = 1$, $2$, $3$, and $\eta_1$ is equal to the value of $\delta$ in Table~\ref{tab:NuFitTable}.}
\label{tab:SterileTable}
\end{table}
We will be interested in the oscillation channels $P_{\mu \mu}$ and $P_{\mu e}$ (and their $CP$ conjugates) for this work. While $P_{\mu\mu}$ is sensitive predominantly to the value of $\sin^2\phi_{24}$, $P_{\mu e}$ is most sensitive to the parameter $\sin^2(2\phi_{e\mu}) \equiv \sin^2(2\phi_{14})\sin^2\phi_{24} = 4|U_{e4}|^2 |U_{\mu 4}|^2$. This is the free parameter seen most often in sterile neutrino searches at short baselines, measuring $P_{\mu e}$ or $P_{e\mu}$. Constraints on the remaining parameter space come from reactor neutrino experiments measuring $P_{ee}$ and $P_{\bar{e}\bar{e}}$, sensitive to $\sin^2\phi_{14}$.

\subsection{Non-standard Neutrino Interactions (NSI)}
\label{subsec:NSI}
We consider the following dimension-six four fermion operator mediating non-standard neutrino interactions:
\begin{equation}
\mathcal{L}_\text{NSI} = -2\sqrt{2}G_F \left( \bar{\nu}_\alpha \gamma_\rho \nu_\beta \right) \left( \epsilon_{\alpha\beta}^{f\tilde{f} L} \bar{f}_L \gamma^\rho \tilde{f}_L + \epsilon_{\alpha\beta}^{f\tilde{f} R} \bar{f}_R \gamma^\rho \tilde{f}_R \right) + \text{h.c.},
\end{equation}
where $G_F$ is the Fermi constant and $\epsilon_{\alpha\beta}$ represent the strength, relative to the weak interactions, of NSI between neutrinos of flavor $\alpha$ and $\beta$ with fermions $f$ and $\tilde{f}$ of chirality $s$. As is standard (see, e.g., Refs.~\cite{deGouvea:2015ndi,Friedland:2004pp,Friedland:2004ah,Friedland:2005vy,Yasuda:2010hw,GonzalezGarcia:2011my,Choubey:2014iia,Friedland:2006pi,Ohlsson:2012kf,Kikuchi:2008vq}), we make several assumptions:
\begin{itemize}
\item $f = \tilde{f} = e$, $u$, $d$ -- we only consider diagonal, neutral current interactions with charged, first-generation fermions.
\item We only consider NSI effects during propagation. For a recent investigation of source, detector, and propagation effects in a long-baseline context, see Ref.~\cite{Blennow:2016etl}.
\item For propagation through earth, we define $\epsilon_{\alpha\beta} \equiv \sum_f \epsilon^f_{\alpha\beta} n_f/n_e$, with $\epsilon_{\alpha\beta}^f \equiv \epsilon_{\alpha\beta}^{f f L} + \epsilon_{\alpha\beta}^{f f R}$ and $n_f$ the number density of fermion $f$. We also assume that $n_u = n_d = 3n_e$.
\end{itemize}

With NSI, the interaction potential for a constant-density region is modified:
\begin{equation}
V_{\alpha\beta} \longrightarrow A \left(\begin{array}{c c c} 1 + \epsilon_{ee} & \epsilon_{e\mu} & \epsilon_{e\tau} \\ \epsilon_{e\mu}^* & \epsilon_{\mu\mu} & \epsilon_{\mu\tau} \\ \epsilon_{e\tau}^* & \epsilon_{\mu\tau}^* & \epsilon_{\tau\tau}\end{array}\right).
\end{equation}
In general, the addition of NSI amounts to nine new parameters, as the off-diagonal elements of $V_{\alpha\beta}$ are complex. Since one element proportional to the identity may be absorbed as a phase in oscillations, we redefine $V'_{\alpha\beta} = V_{\alpha\beta} - \epsilon_{\mu\mu}\mathbbm{1}$. When considering antineutrino oscillations, $A\to -A$ (as in the three-neutrino hypothesis) and $\epsilon_{\alpha\beta} \to \epsilon_{\alpha\beta}^*$.

As with the sterile neutrino hypothesis, we give a set of illustrative NSI parameters for comparison against the three-neutrino hypothesis in figures.
\begin{table}[!htbp]
\begin{center}
\begin{tabular}{|c||c|c|c|c|c|c|}
\hline
Parameter & $\epsilon_{ee}$ & $\epsilon_{e\mu}$ & $\epsilon_{e\tau}$ & $\epsilon_{\mu\mu}$ & $\epsilon_{\mu\tau}$ & $\epsilon_{\tau\tau}$ \\ \hline
Value & $0$ & $0.5e^{i\pi/3}$ & $0.5e^{-i\pi/4}$ & 0$^\star$ & 0 & $-1$ \\ \hline
\end{tabular}
\end{center}
\caption{Input values used for an illustrative NSI hypothesis for comparisons in figures throughout this work. Three-neutrino parameters are equal to their values in Table~\ref{tab:NuFitTable}. We include a star on the value of $\epsilon_{\mu\mu}$ as a reminder that this parameter is set to zero in our analysis, as discussed in the text.}
\label{tab:NSITable}
\end{table}
For a thorough discussion of the bounds on NSI parameters for neutrino propagation through the earth, we refer the reader to Refs.~\cite{Biggio:2009nt,Ohlsson:2012kf,Gonzalez-Garcia:2013usa}.

\section{The Hyper-Kamiokande Experiment}
\label{sec:HK}

The Hyper-Kamiokande (Hyper-K) Experiment is a proposed next-generation neutrino experiment that utilizes two water Cerenkov detectors with total mass of 0.99 Megatons (0.56 Mton fiducial) located in the Tochibora Mine, 8km south of the existing Super-Kamiokande (Super-K) experiment~\cite{Abe:2015zbg}. The upgraded Japan Proton Accelerator Research Complex (J-PARC) proton synchrotron beam is expected to deliver $1.56\times 10^{22}$ protons on target over ten years of data collection. In Section~\ref{subsec:BeamCap}, we discuss the capability of Hyper-K using the neutrino beam originating at J-PARC, 295 km away from the detector, and in Section~\ref{subsec:AtmosCap}, we discuss the capability of Hyper-K in utilizing atmospheric neutrinos. A recent proposal (see Ref.~\cite{Abe:2016ero}) suggests placing one detector in Korea for a longer baseline, however we consider only the original proposal. Refs.~\cite{Fukasawa:2016lew,Liao:2016orc,Ghosh:2017ged} discuss the potential of this setup in light of NSI.

\subsection{Beam-based detector capabilities}
\label{subsec:BeamCap}
The J-PARC beam is capable of operating in two modes, neutrino and antineutrino, in which the dominant contributions to the beam are $\nu_\mu$ and $\bar{\nu}_\mu$, respectively. Ref.~\cite{Abe:2015zbg} has determined that the optimal ratio for operating in these two modes is $1:3$ for $\nu :\bar{\nu}$, and so we take this, and an assumption of ten years of data collection, for our analysis. The two analyses performed are the appearance ($\nu_\mu \to \nu_e$) and disappearance ($\nu_\mu \to \nu_\mu$) channels. Both channels assume bins of $50$ MeV, and we smear\footnote{This smearing and our attempted replication of reconstruction efficiencies lead to apparent discrepancies between our simulation and that of Ref.~\cite{Abe:2015zbg}, where our distributions appear more smeared, particularly in the disappearance channels. We find that changing the smearing has little-to-no impact on the results of this work, as long as signal and background rates are normalized to those presented in Ref.~\cite{Abe:2015zbg}.} the reconstructed energy distributions attempting to match the results of Ref.~\cite{Abe:2015zbg}. Electron (appearance) candidates range in energy between $100$ MeV and $1.25$ GeV, where muon (disappearance) candidates range between $200$ MeV and $10$ GeV. Using projected fluxes from Ref.~\cite{Abe:2015zbg}, neutrino-nucleon cross sections from Ref.~\cite{Formaggio:2013kya}, and oscillation probabilities calculated given a particular hypothesis, we determine the expected event yield at Hyper-K assuming ten years of data collection with a ratio of $1:3$ for $\nu :\bar{\nu}$ modes.
\begin{figure}[!htbp]
\centering
\includegraphics[width=0.9\linewidth]{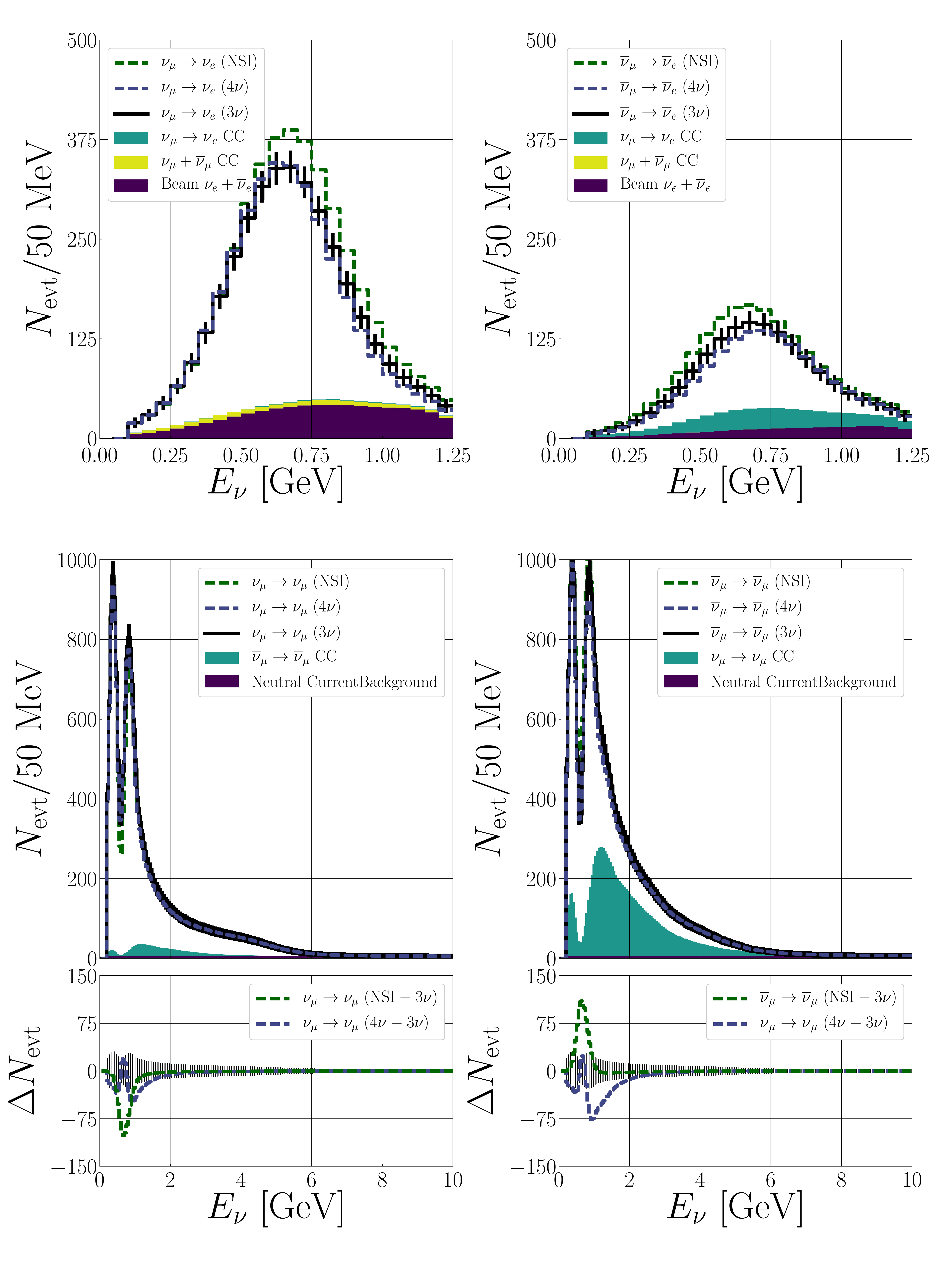}
\caption{Expected yields in the appearance channels (top) and disappearance channels (middle) assuming ten years of data collection at the Hyper-Kamiokande experiment with a ratio of $1:3$ for the duration of neutrino and antineutrino modes. The left panels display yields during neutrino mode, and the right panels display yields during antineutrino mode. In each panel, backgrounds are displayed as a stacked histogram, with opposite-sign signal events shown in teal, muon-neutrino misidentification shown in yellow in the appearance channels, beam contamination backgrounds in purple in the appearance channels, and neutral current backgrounds in purple in the disappearance channels. As discussed in the text, expected neutral current event rates have been included by inflating the beam contamination background for the appearance channels, and as a flat background in the disappearance channels. Three different sets of overall signal plus background yields are shown in each panel: for a three-neutrino scenario assuming parameters from Table~\ref{tab:NuFitTable} (black, including statistical error bars), a four-neutrino scenario assuming parameters from Table~\ref{tab:SterileTable} (blue, dashed), and a non-standard interaction scenario assuming parameters from Table~\ref{tab:NSITable} (green, dashed). The bottom panels additionally show differences in the number of expected events per bin between the NSI and three-neutrino scenarios (green) and four- and three-neutrino scenarios (blue).}
\label{fig:EvtYields}
\end{figure}

Fig.~\ref{fig:EvtYields} displays expected event yields at Hyper-K assuming ten years of data collection. The top panels display appearance channels for $\nu$ mode (left) and $\bar{\nu}$ mode (right), and the bottom panels display disappearance channels for $\nu$ mode (left) and $\bar{\nu}$ mode (right). For appearance channels, we consider background contributions due to opposite sign signal (``$\bar{\nu}_\mu \to \bar{\nu}_e$ CC'' and ``$\nu_\mu \to \nu_e$ CC'', teal), unoscillated muon contamination (``$\nu_\mu + \bar{\nu}_\mu$ CC'', yellow), and unoscillated electron contamination (``Beam $\nu_e + \bar{\nu}_e$'', purple). As we do not have strong information regarding the neutral current backgrounds, we have inflated the unoscillated electron contamination to match background rates in Ref.~\cite{Abe:2015zbg}. For disappearance channels, we include opposite sign signal (``$\bar{\nu}_\mu \to \bar{\nu}_\mu$ CC'' and ``$\nu_\mu\to\nu_\mu$ CC'', teal) and a flat neutral current background (purple). For each panel, we display total yields assuming three neutrinos exist (using the parameters in Table~\ref{tab:NuFitTable}, black, with statistical error bars shown), assuming four neutrinos exist (using the illustrative case in Table~\ref{tab:SterileTable}, blue), and assuming NSI (using the illustrative case in Table~\ref{tab:NSITable}, green).

\subsection{Atmospheric-based detector capabilities}
\label{subsec:AtmosCap}
\begin{figure}[!htbp]
\begin{center}
\includegraphics[width=0.7\linewidth]{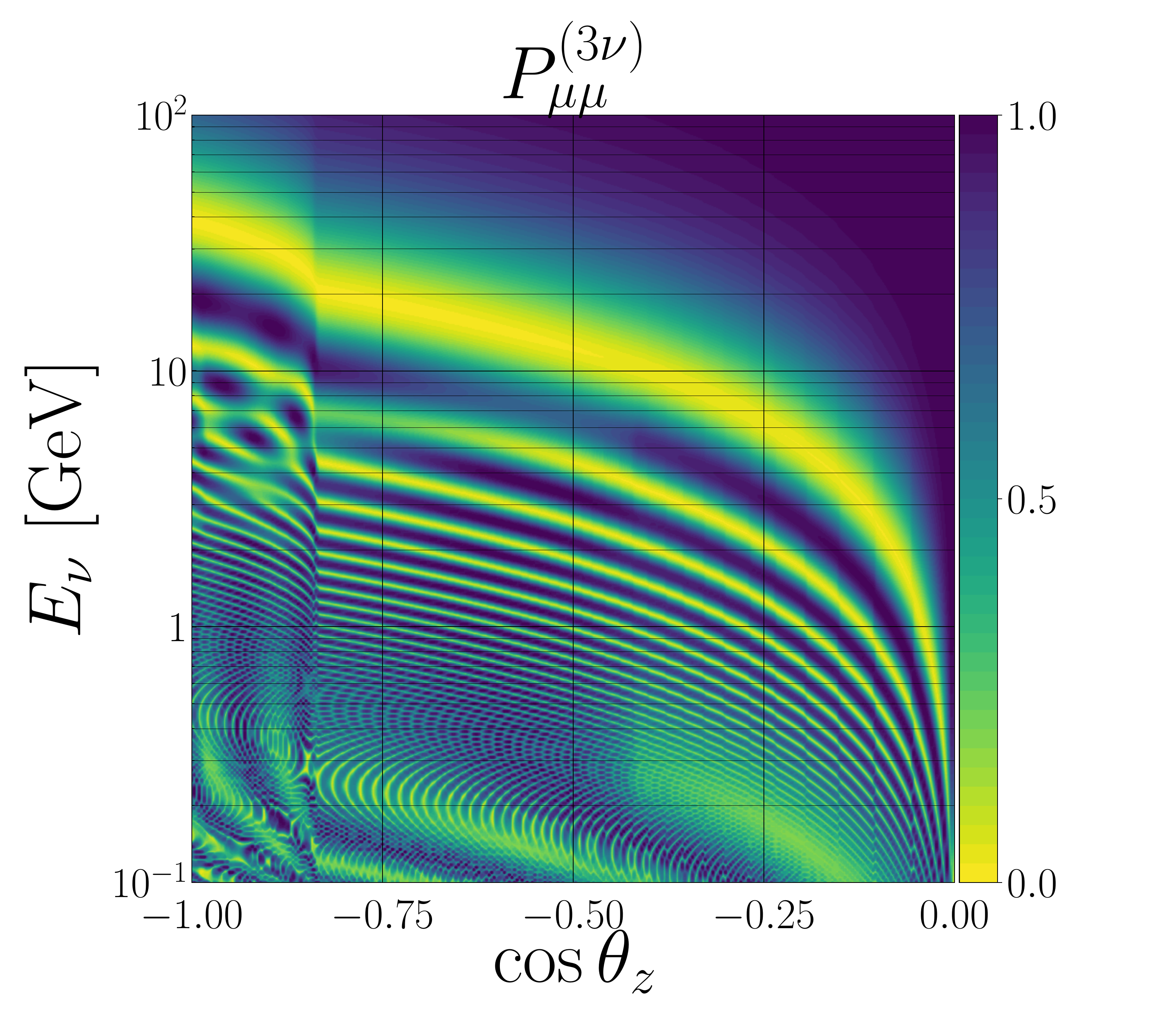}
\end{center}
\caption{Oscillation plot of $P_{\mu\mu}$ assuming three neutrinos exist with parameters in Table~\ref{tab:NuFitTable} as a function of $\cos{\theta_z}$, where $\theta_z$ is the zenith angle (with $\theta_z = 0$ being directly overhead) and neutrino energy $E_\nu$. Here, we assume a piecewise-constant density profile for the Earth so that Eq.~(\ref{eq:AtmosProb}) may be utilized. For comparison against previous works (see, e.g., Ref.~\cite{Abe:2014gda}), we display only $\cos{\theta_z} \in [-1, 0]$, and find that our results match those using a more sophisticated density profile like PREM~\cite{Dziewonski:1981xy}. We calculate probabilities over the entire range of $\theta_z$ in practice.}
\label{fig:Oscillog3Nu}
\end{figure}
\begin{figure}[!htbp]
\begin{center}
\includegraphics[width=\linewidth]{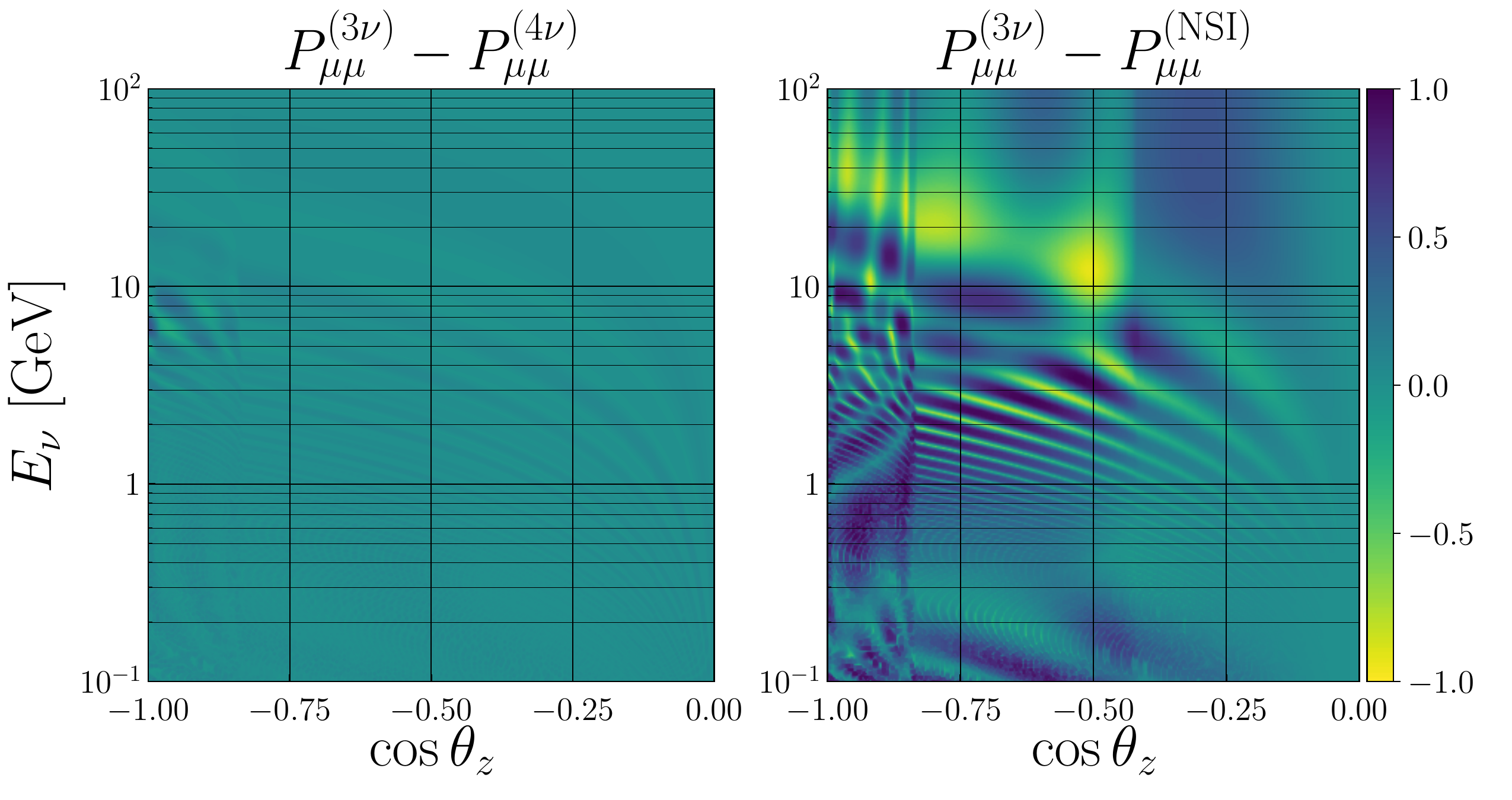}
\end{center}
\caption{Differences in the oscillation probability $P_{\mu\mu}$ with the three neutrino case shown in Fig.~\ref{fig:Oscillog3Nu} for a four-neutrino scenario with parameters from Table~\ref{tab:SterileTable} (left) and an NSI scenario with parameters from Table~\ref{tab:NSITable} (right).}
\label{fig:OscillogDiffs}
\end{figure}
In addition to neutrinos produced by the J-PARC beam, Hyper-K is sensitive to atmospheric neutrinos, similar to its predecessor Super-K. The dominant channel contributing to atmospheric neutrino oscillations at Hyper-K is $P_{\mu\mu}$. Fig.~\ref{fig:Oscillog3Nu} displays an oscillogram of $P_{\mu\mu}$ for a three-neutrino case as a function of the (cosine of the) zenith angle and neutrino energy. Additionally, we show differences in oscillation probability in Fig.~\ref{fig:OscillogDiffs} between a three-neutrino case and an four-neutrino case (left) and between a three-neutrino case and an NSI case (right). While the figures here only display the range $\cos{\theta_z} \in [-1, 0]$ (upward-going neutrinos) for the sake of comparison, the entire range of zenith angles is calculated in practice. Despite using a piecewise-constant density profile, the behavior here matches that seen in Ref.~\cite{Abe:2014gda}.

\begin{figure}[!htbp]
\centering
\includegraphics[width=0.8\linewidth]{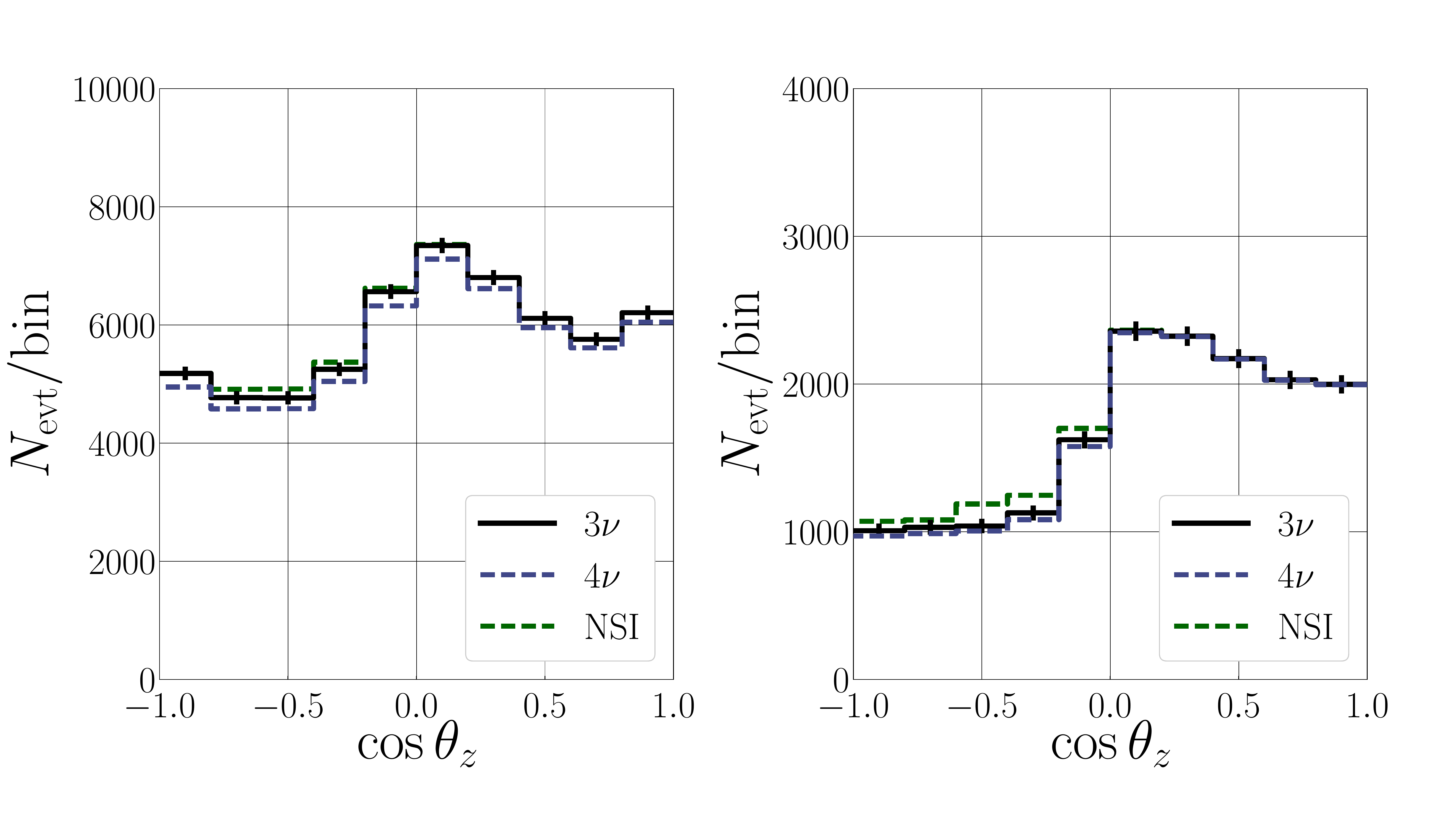}
\caption{Expected muon-type event yields at Hyper-Kamiokande assuming a three-neutrino scenario with parameters from Table~\ref{tab:NuFitTable} (black, including statistical error bars), assuming a four-neutrino scenario with parameters from Table~\ref{tab:SterileTable} (blue, dashed), and assuming NSI exist with parameters from Table~\ref{tab:NSITable} (green, dashed). Oscillation probabilities are calculated as discussed in the text, and then convolved with fluxes from Ref.~\cite{Honda:2015fha}. The event distribution is divided into low-energy ($E_\nu < 1.3$ GeV, left) and high-energy ($E_\nu > 1.3$ GeV, right) samples, and smeared by $10^\circ$ ($5^\circ$) for the low- (high-) energy distribution due to the correlation between the incident muon neutrino and outgoing muon tracks. Distributions are then binned in ten bins of $\cos{\theta_z}$, as seen in the figure.}
\label{fig:AtmosYields}
\end{figure}

Ref.~\cite{Honda:2015fha} details the expected atmospheric neutrino flux at the location of Super-K, and we estimate the yield after ten years at Hyper-K by increasing the Super-K exposure by a factor of $20$. We only consider measurements of muon-type neutrinos in the detector -- this relies on the muon (anti)neutrino flux in the upper atmosphere multiplied by $P_{\mu\mu}$ ($P_{\bar{\mu}\bar{\mu}}$) and the electron (anti)neutrino flux multiplied by $P_{e\mu}$ ($P_{\bar{e}\bar{\mu}}$). Considering appearance of electron- and tau-type neutrinos would improve results by measuring the oscillation probabilities $P_{\mu e}$, $P_{\mu\tau}$, etc., however we analyze only muon-type neutrino measurements for simplicity. As with Super-K, we divide up muon neutrino samples into sub-GeV ($E_\nu < 1.3$ GeV) and multi-GeV events, and we divide up the incoming direction of the neutrinos (the zenith angle $\theta_z$) into ten bins of $\cos\theta_z$. Additionally, we smear the reconstructed low- (high-) energy distribution by $10^\circ$ ($5^\circ$) given the correlation between the incident muon neutrino and outgoing muon track. Expected event counts as a function of $\cos{\theta_z}$ after smearing and binning are shown in Fig.~\ref{fig:AtmosYields}. Comparing Figs.~\ref{fig:OscillogDiffs} and \ref{fig:AtmosYields}, we see that, for the majority of energies, $P_{\mu\mu}^{(4\nu)} < P_{\mu\mu}^{(3\nu)}$, leading to fewer expected events in Fig.~\ref{fig:AtmosYields}. Also, we see that, predominantly for higher energy neutrinos ($E_\nu \gtrsim 1$ GeV), $P_{\mu\mu}^{(\mathrm{NSI})} > P_{\mu\mu}^{(3\nu)}$, leading to a higher number of expected events in the right panel of Fig.~\ref{fig:AtmosYields}.

\subsection{Analysis method}
\label{subsec:Analysis}
Our analysis method is as follows. First, we simulate expected yields for beam-based and atmospheric neutrino detection assuming three neutrinos exist, with parameters shown in Table~\ref{tab:NuFitTable}. Then, given a test hypothesis with parameters\footnote{For the sterile neutrino hypothesis, we use the parameter space $\vec{\vartheta} =$ ($\phi_{12},$ $\phi_{13},$ $\phi_{23},$ $\Delta m_{12}^2,$ $\Delta m_{13}^2,$ $\eta_1,$ $\sin^2\phi_{24},$ $4|U_{e4}|^2|U_{\mu 4}|^4,$ $\sin^2\phi_{34},$ $\eta_2,$ $\eta_3,$ $\Delta m_{14}^2$), where we use $4|U_{e4}|^2|U_{\mu 4}|^2 = 4\sin^2\phi_{14}\cos^2\phi_{14}\sin^2\phi_{24}$ as an independent parameter to compare against short-baseline sterile neutrino searches.} $\vec{\vartheta}$, we calculate a chi-squared function. Included in the chi-squared function are Gaussian priors on the solar mass splitting\footnote{The one-sigma range on $\Delta m_{12}^2$ in Table~\ref{tab:NuFitTable} is nearly symmetric -- we approximate the one-sigma range to be $\Delta m_{12}^2 \ (7.50 \pm 0.18) \times 10^{-5}$ eV$^2$ in our analysis.} $\Delta m_{12}^2$ and $|U_{e2}|^2$, where the one-sigma ranges are given in Table~\ref{tab:NuFitTable}. We also include normalization uncertainties in the chi-squared function: 5\% signal and background uncertainties for the beam-based data and 10\% for the atmospheric-based data. While certain parameters ($\sin^2{\phi_{34}}, \eta_{2,3}$ for the sterile neutrino hypothesis and $\epsilon_{ee}$ and $\epsilon_{\mu\tau}$ for the NSI hypothesis) were set to zero for the illustrative examples listed in Tables~\ref{tab:SterileTable} and \ref{tab:NSITable}, none of the parameters (except $\epsilon_{\mu\mu}$ as discussed above) are fixed in our analysis. This amounts to 12 free parameters for the sterile neutrino scenario and 14 for the NSI scenario.

We then use the Markov Chain Monte Carlo package {\sc emcee} to calculate posterior likelihood distributions in the parameter space of a particular test hypothesis, and from these, we calculate one- and two-dimensional chi-squared distributions, marginalized over all other parameters~\cite{ForemanMackey:2012ig}. We define the $95\%$ ($99\%$) CL sensitivity reach of Hyper-K as regions where $\chi^2 - \chi^2_\text{min} > 5.99$ ($9.21$) for two-dimensional figures and $\chi^2 - \chi^2_\text{min} > 3.84$ ($6.63$) for one-dimensional figures. For each new physics hypothesis, we perform this analysis using only beam-based results, and using a combination of beam- and atmospheric-based results.

\section{Results}
\label{sec:Results}

\subsection{Sterile Neutrino}
\begin{figure}[!htbp]
\centering
\includegraphics[width=0.8\linewidth]{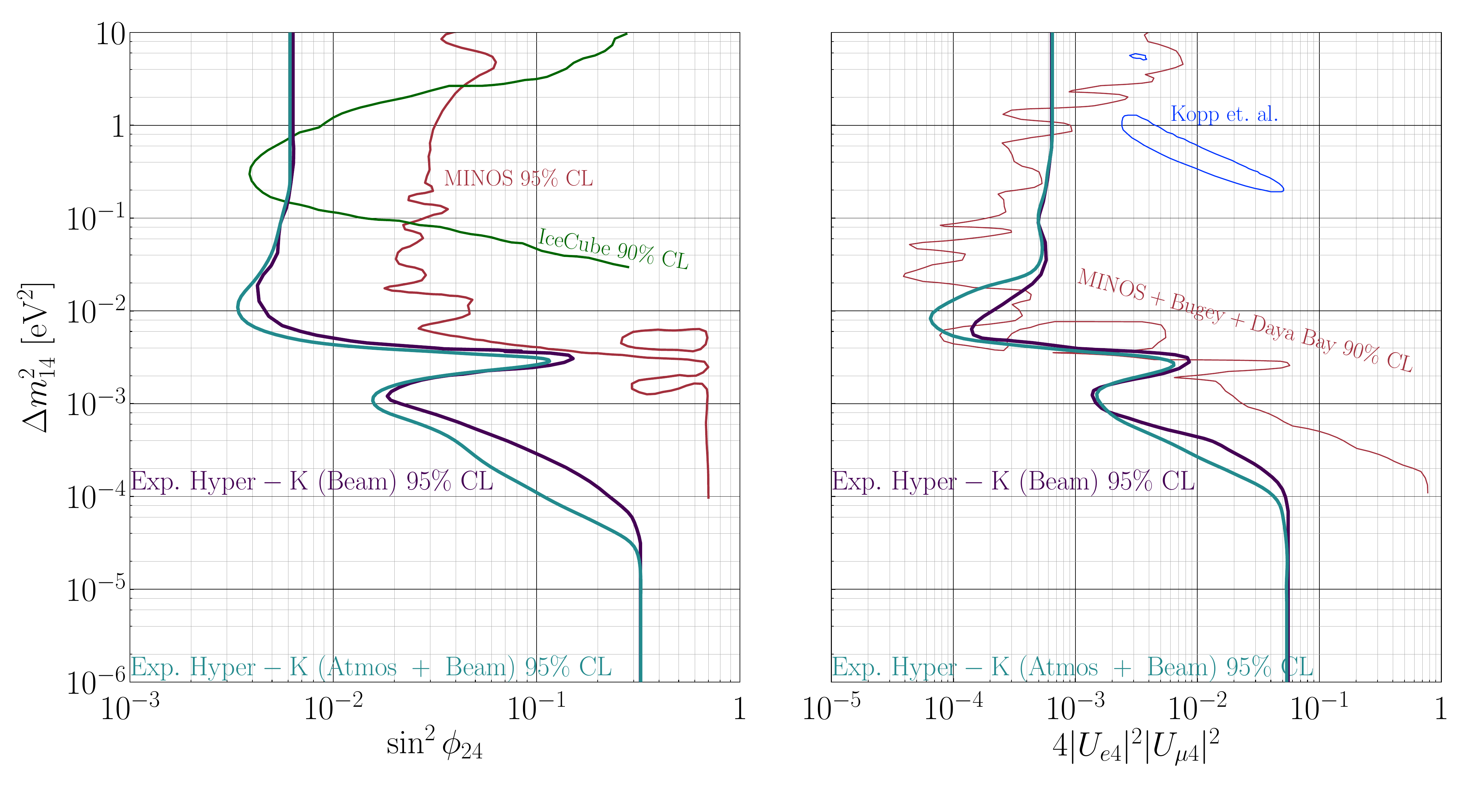}
\caption{Expected sensitivity to a fourth neutrino assuming ten years of only beam-based data at Hyper-Kamiokande at 95\% CL (purple) and including atmospheric-based data (teal). Regions above and to the right of these curves will be excluded at 95\% CL by Hyper-K if only three neutrinos exist. The left panel displays sensitivity in the $\sin^2\phi_{24}$ - $\Delta m_{14}^2$ plane, with contributions predominantly from the disappearance channels, and the right panel displays sensitivity in the $4|U_{e 4}|^2 |U_{\mu 4}|^2 = 4\sin^2\phi_{14}\sin^2\phi_{24}\cos^2\phi_{14}$ - $\Delta m_{14}^2$ plane, with contributions predominantly from the appearance channels. All unseen parameters are marginalized in each panel. In the left panel, we display existing bounds from the MINOS~\cite{MINOS:2016viw} (red, 95\% CL) and IceCube~\cite{TheIceCube:2016oqi} (green, 90\% CL) experiments. In the right panel, we display the most competitive existing bound in this parameter space, a combined analysis from the MINOS, Bugey, and Daya Bay experiments~\cite{Adamson:2016jku} (red, 90\% CL) and the preferred parameter space of various reactor and short-baseline sterile neutrino hints from a combined global analysis in Ref.~\cite{Kopp:2013vaa} (blue). Gaussian priors are included on the values of $|U_{e2}|^2 = 0.2994 \pm 0.0117$ and $\Delta m_{12}^2 = (7.50 \pm 0.18) \times 10^{-5}$ eV$^2$. Estimated sensitives are calculated utilizing {\sc emcee}~\cite{ForemanMackey:2012ig}.}
\label{fig:Sterile_Excl}
\end{figure}

Here we generate data consistent with only three neutrinos existing, and analyze the sensitivity of the Hyper-K experiment to detect a fourth neutrino. Fig.~\ref{fig:Sterile_Excl} displays the sensitivity reach of the Hyper-K experiment in the $\sin^2\phi_{24}$ - $\Delta m_{14}^2$ (left) and $4|U_{e4}|^2 |U_{\mu 4}|^2$ - $\Delta m_{14}^2$ (right) planes using only data from the beam-based capabilities (purple). The region above and to the right of each curve will be excluded at 95\% CL by Hyper-K if only three neutrinos exist. In both panels, we see that in the high-$\Delta m_{14}^2$ range, oscillations average out, and in the low-$\Delta m_{14}^2$ range, while oscillations due to the fourth mass eigenstate are not detectable, non-zero mixing angles $\phi_{i4}$ can impact the unitarity of the $3\times 3$ sub-matrix of the $4\times 4$ PMNS matrix, and may be detectable at Hyper-K. This feature has been discussed in the context of long-baseline neutrino oscillations (at DUNE) previously in Ref.~\cite{Berryman:2015nua}. We also see a feature in both panels of Fig.~\ref{fig:Sterile_Excl} where sensitivity is weaker for $\Delta m_{14}^2 \sim 10^{-3} - 10^{-2}$ eV$^2$. This comes from the fact that $\Delta m_{13}^2$ is in this range, and there is degeneracy between the mixing angles $\phi_{i4}$ and $\phi_{i3}$.

Fig.~\ref{fig:Sterile_Excl} additionally displays results of our analysis incorporating both beam- and atmospheric-based detection (teal). We see small improvement in both the $\sin^2\phi_{24}$ - and $4|U_{e4}|^2|U_{\mu 4}|^2$ - $\Delta m_{14}^2$ planes, however it is limited, likely due to the 10\% normalization uncertainty included in the atmospheric neutrino sample. The Super-Kamiokande collaboration noted that oscillations due to sterile neutrinos average out above $\Delta m_{14}^2 \gtrsim 10^{-1}$ eV$^2$~\cite{Abe:2014gda}, and we see this same behavior in Fig.~\ref{fig:Sterile_Excl}. If a more thorough analysis were performed, particularly including the measurement of electron-type neutrinos in the atmospheric data sample, there would likely be improvement, particularly in the right panel from the sensitivity to two additional oscillation probability channels -- $P_{\mu e}$ and $P_{ee}$.

\subsection{Non-standard Neutrino Interactions}
\begin{figure}[!htbp]
\centering
\includegraphics[width=0.7\linewidth]{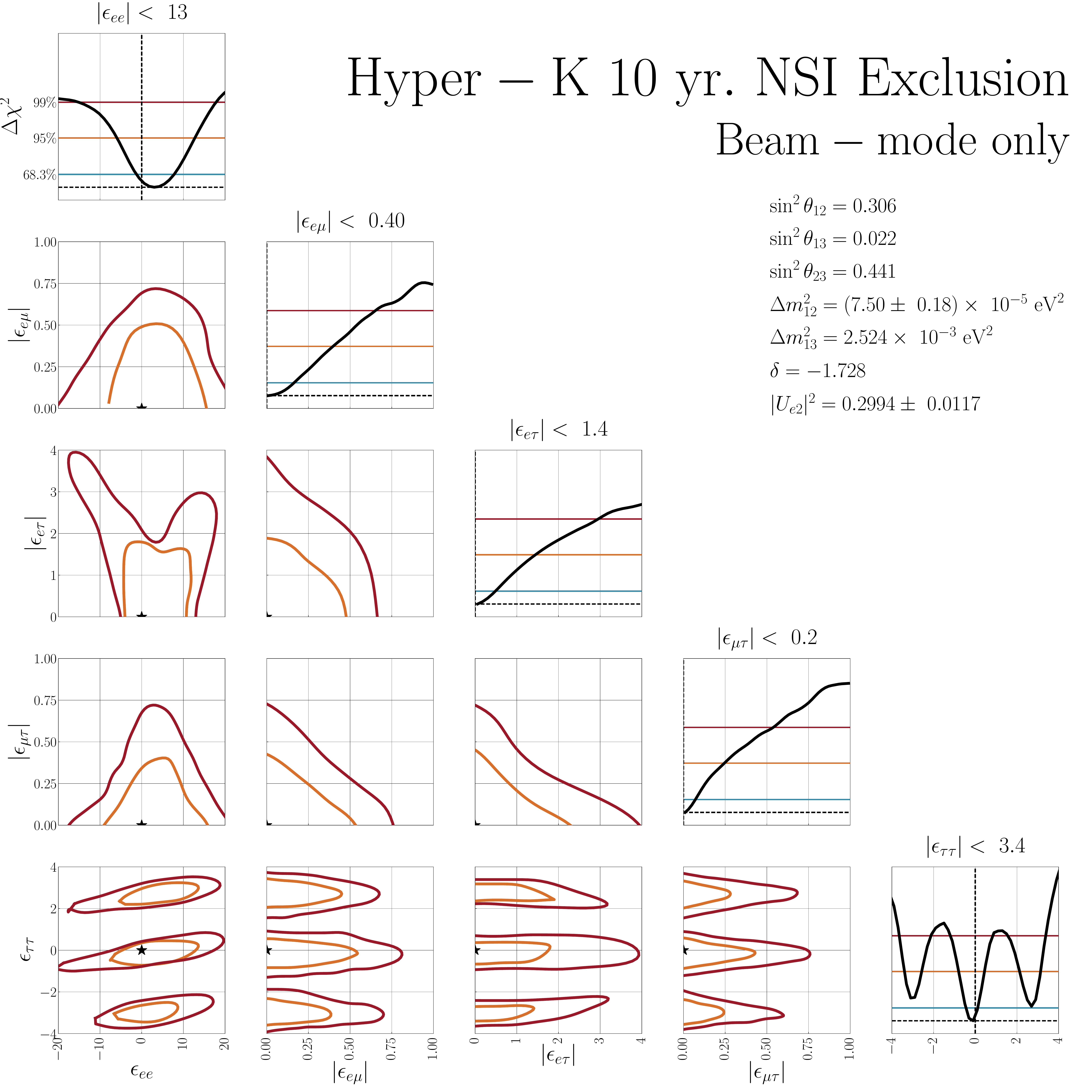}
\caption{Expected sensitivity to non-zero NSI assuming ten years of beam-based data collection at Hyper-Kamiokande at 95\% CL (orange) and 99\% CL (red). In each panel, all unseen parameters, including three-neutrino parameters and phases of off-diagonal NSI, are marginalized. The top panel of each column displays expected one-dimensional $\Delta \chi^2$ sensitivity for each parameter, including horizontal lines displaying $68.3\%$ (blue), $95\%$ (orange), and $99\%$ (red) CL. Above each column, the expected limits at $95\%$ CL for each parameter are shown. Gaussian priors are included on the values of $|U_{e2}|^2 = 0.2994 \pm 0.0117$ and $\Delta m_{12}^2 = (7.50 \pm 0.18) \times 10^{-5}$ eV$^2$. Estimated sensitivities are calculated utilizing {\sc emcee}~\cite{ForemanMackey:2012ig}.}
\label{fig:NSI_Beam_Excl}
\end{figure}

Fig.~\ref{fig:NSI_Beam_Excl} displays the expected sensitivity at 95\% (orange) and 99\% (red) CL to non-standard neutrino interactions assuming ten years of beam-based data collection at Hyper-Kamiokande. In each panel, all unseen parameters (including three-neutrino parameters and phases of complex NSI) are marginalized. At the top of each column, a one dimensional $\Delta \chi^2$ plot is shown for each parameter, including horizontal lines corresponding to 68.3\% (blue), 95\% (orange), and 99\% (red) CL.
We note several degeneracies throughout this figure: most notable are the features in the $\epsilon_{ee}$ - $|\epsilon_{e\tau}|$ plane and the degeneracy between $\epsilon_{\tau\tau} = 0$ and $\epsilon_{\tau\tau} \simeq \pm 3$. Degeneracies of this nature have been discussed in the context of long-baseline oscillations in Refs.~\cite{Friedland:2006pi,Kopp:2010qt,GonzalezGarcia:2011my,Coloma:2011rq,deGouvea:2015ndi,Coloma:2015kiu,Liao:2016hsa,Bakhti:2016prn,Coloma:2016gei,deGouvea:2016pom,Blennow:2016etl,Blennow:2016jkn,Fukasawa:2016lew,Ghosh:2017ged}. The $\epsilon_{\tau\tau}$ degeneracy has been discussed at length in Ref.~\cite{deGouvea:2015ndi}, and it arises from a degeneracy between $\epsilon_{\tau\tau}$ and $\theta_{23}$ for a non-maximal physical value of $\theta_{23}$ as we have here ($\sin^2\theta_{23} = 0.441$).

\begin{figure}[!htbp]
\centering
\includegraphics[width=0.7\linewidth]{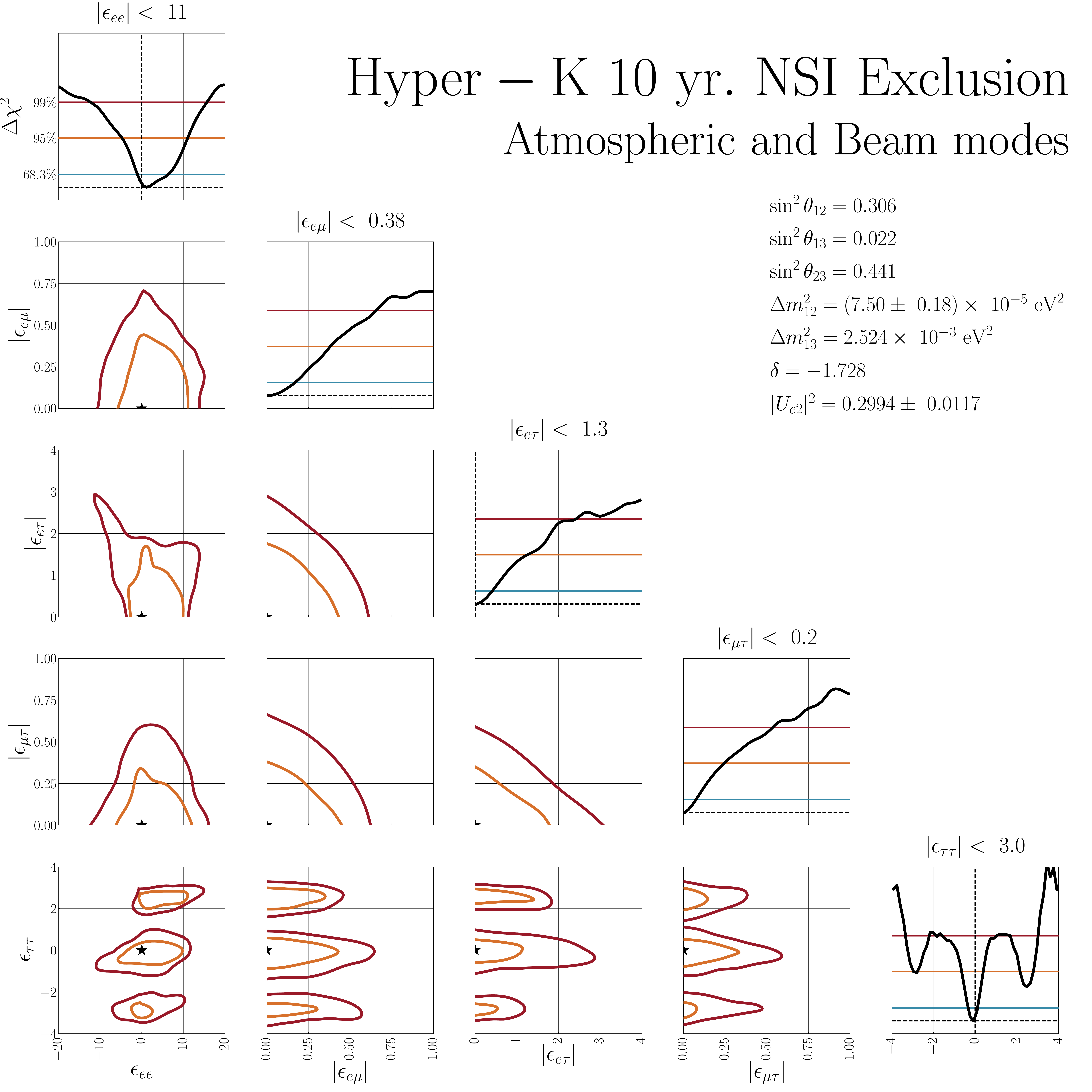}
\caption{Expected sensitivity to non-zero NSI assuming ten years of beam- and atmospheric-based data collection at Hyper-Kamiokande at 95\% CL (orange) and 99\% CL (red). In each panel, all unseen parameters, including three-neutrino parameters and phases of off-diagonal NSI, are marginalized. The top panel of each column displays expected one-dimensional $\Delta \chi^2$ sensitivity for each parameter, including horizontal lines displaying $68.3\%$ (blue), $95\%$ (orange), and $99\%$ (red) CL. Above each column, the expected limits at $95\%$ CL for each parameter are shown. Gaussian priors are included on the values of $|U_{e2}|^2 = 0.2994 \pm 0.0117$ and $\Delta m_{12}^2 = (7.50 \pm 0.18) \times 10^{-5}$ eV$^2$. Estimated sensitivities are calculated utilizing {\sc emcee}~\cite{ForemanMackey:2012ig}.}
\label{fig:NSI_AtmosBeam_Excl}
\end{figure}

\begin{figure}[!htbp]
  \begin{center}
    \includegraphics[width=0.35\textwidth]{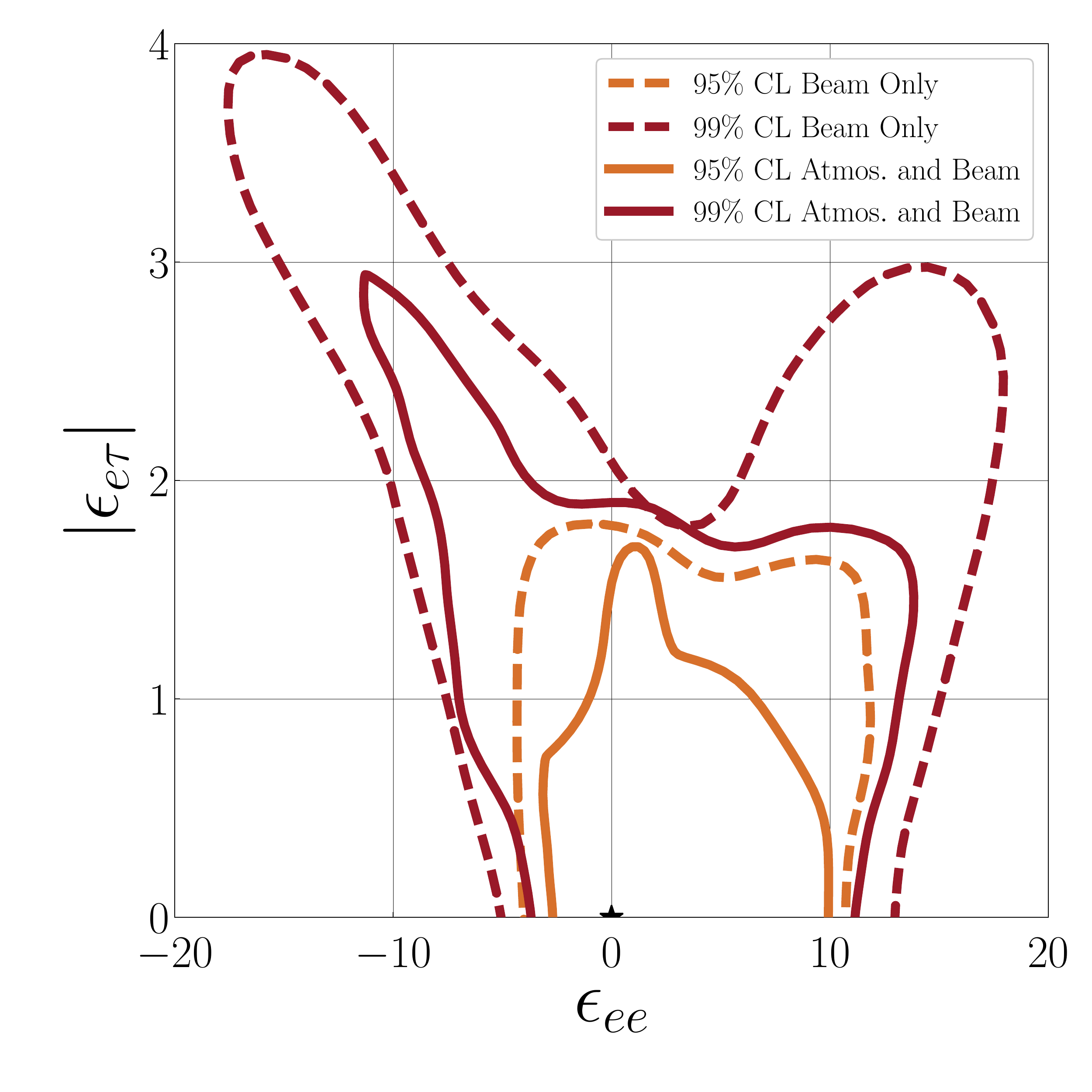}
  \end{center}
  \caption{Improvement in sensitivity to NSI parameters at Hyper-K between considering only beam-based data (dashed lines) and including atmospheric-based data as well (solid lines). Contours shown are 95\% CL (orange) and 99\% CL (red) in the $\epsilon_{ee}$ - $|\epsilon_{e\tau}|$ plane -- all unseen parameters, including the phase on $\epsilon_{e\tau}$ are marginalized. Sensitivities are estimated using {\sc emcee}~\cite{ForemanMackey:2012ig}.}
  \label{fig:Improvement}
\end{figure}

Results of the analysis including both beam- and atmospheric-based data are shown in Fig.~\ref{fig:NSI_AtmosBeam_Excl}. A direct comparison between this and the results of Super-K~\cite{Mitsuka:2011ty} and IceCube~\cite{Day:2016shw,Salvado:2016uqu} is non-trivial, as our analysis includes all NSI parameters simultaneously, as well as allowing for the off-diagonal NSI parameters to be complex and $CP$-violating. Allowing for complex $\epsilon_{\mu\tau}$ has been shown to decrease sensitivity significantly in, e.g., Refs.~\cite{Kopp:2010qt,Coloma:2011rq,deGouvea:2015ndi,Coloma:2015kiu,Blennow:2016jkn}.
While there is not drastic improvement between the results in Fig.~\ref{fig:NSI_Beam_Excl} and Fig.~\ref{fig:NSI_AtmosBeam_Excl}, we note that there is improvement in the degeneracies seen in the $\epsilon_{ee}$ - $|\epsilon_{e\tau}|$ plane as well as in alleviating some of the degeneracy seen for $\epsilon_{\tau\tau}$. For direct comparison of the improvement in the $\epsilon_{ee}$ - $|\epsilon_{e\tau}|$ plane, we show both expected sensitivites in Fig.~\ref{fig:Improvement}.

Comparing the results in Figs.~\ref{fig:NSI_Beam_Excl} and \ref{fig:NSI_AtmosBeam_Excl} with those from a multi-parameter study at DUNE (see Refs.~\cite{deGouvea:2015ndi} and \cite{Coloma:2015kiu}), we see that, even with atmospheric neutrino data, the expected sensitivity reach to NSI at Hyper-K is between a factor of five to ten weaker than that at DUNE. This is unsurprising: NSI effects grow at larger baselines if the same $L/E_\nu$ ratio is being probed -- the baseline length of Hyper-K (295 km) is significantly shorter than that of DUNE (1300 km). A combined analysis could prove useful -- while Hyper-K does not constrain the NSI parameters significantly better than DUNE, the combination of beam- and atmospheric-based data clears up degeneracies that trouble DUNE. With DUNE and Hyper-K data measuring neutrino oscillations in the same range of $L/E_\nu$ values and at vastly different baseline lengths, many of these degeneracies may be lifted with a combination of data. Additionally, as noted in the context of sterile neutrinos, the addition of electron neutrino measurements in the atmospheric-based data would aide in improving NSI sensitivity at Hyper-K, particularly in the parameters $\epsilon_{ee}$, $|\epsilon_{e\mu}|$, and $|\epsilon_{e\tau}|$, which are more relevant for oscillation probabilities $P_{\mu e}$ and $P_{ee}$ than for $P_{\mu\mu}$ and $P_{e\mu}$.

\section{Discussion and Conclusions}
\label{sec:Conclusions}

Upcoming long-baseline, large-statistics neutrino oscillation experiments such as Hyper-Kamiokande and the Deep Underground Neutrino Experiment will be able to measure the remaining parameters regarding three-neutrino mixing and oscillation, and will additionally start to probe whether the mixing is $CP$-invariant. These upcoming experiments will also have the ability to detect physics beyond the three-massive-neutrinos paradigm. In this work, we explored the capability of Hyper-K to detect two of these new-physics hypotheses: the existence of a fourth, sterile neutrino, and the existence of additional neutrino interactions other than the weak interactions.

We discussed the ways in which these new-physics hypotheses manifest themselves in neutrino oscillations at long-baselines, as well as in oscillations of atmospheric neutrinos propagating through the Earth. The latter is important, as the measurement of atmospheric neutrinos is key in the ability of Hyper-K to achieve its physics goals, in addition to the measurement of beam-based neutrinos from J-PARC. The specifics of the beam- and atmospheric-based neutrino capabilities were discussed in some detail, including discussing backgrounds considered in the beam-based measurements.

We performed simulations assuming the Hyper-K detectors will have a total mass of 0.99 Megatons (0.56 Mton fiducial), and that the experiment will last ten years. While more recent proposals have suggested placing one of the two Hyper-K detectors in Korea, we considered only the proposal that both are in Japan, 295 km from the origin of the neutrino beam at J-PARC. We have assumed that the beam, capable of running in both neutrino and antineutrino modes, has a ratio of runtime of $1:3$ for $\nu : \bar{\nu}$ modes. Given the size of the detector, we estimate that the total yield of atmospheric neutrinos will be $20$ times that of Hyper-K's predecessor, Super-Kamiokande. With conservative estimates on this, zenith angle smearing, and smearing over expected energy, as well as only considering muon-type neutrinos, we calculate the expected yields for low- and high-energy neutrinos at Hyper-K.

The yields we calculate are used, along with conservative estimates for signal and background normalization uncertainties, in a Markov Chain Monte Carlo algorithm to calculate expected sensitivities using a chi-squared statistic approach. We presented our results in terms of sensitivity reach of the Hyper-K experiment at 95\% and 99\% CL, showing both the expected reach for beam-based measurements only, and the improvement when atmospheric-based measurements are included as well. We find that Hyper-K is able to reach new regions of parameter space that have yet to be explored by existing experiments, and that it will be competitive with DUNE. The results shown assumed that the neutrino mass hierarchy is discovered prior to Hyper-K collecting data, and that the hierarchy is normal. We also only included muon-type neutrinos in the atmospheric-based data sample: including electron appearance in this sample would improve sensitivity to new physics as well. 

We also briefly discussed the complementarity of DUNE and Hyper-K, as the two experiments measure neutrino oscillations in the same range of $L/E_\nu$, the baseline length divided by the neutrino energy, however they have vastly different values for $L$ and $E_\nu$. This overlap in $L/E_\nu$ allows the experiments to probe for new physics phenomena in complementary ways, and a combined analysis between the experiments would be able to better search for these new phenomena.

\begin{acknowledgments}
We would like thank Andr\'e de Gouv\^ea and Jeff Berryman for useful discussions regarding this work. This work is supported in part by DOE grant \#de-sc0010143. We also acknowledge the use of the Quest computing cluster at Northwestern University for a portion of this research.
\end{acknowledgments}

\bibliographystyle{apsrev-title}
\bibliography{HKBib}{}

\begin{thebibliography}{113}
\expandafter\ifx\csname natexlab\endcsname\relax\def\natexlab#1{#1}\fi
\expandafter\ifx\csname bibnamefont\endcsname\relax
  \def\bibnamefont#1{#1}\fi
\expandafter\ifx\csname bibfnamefont\endcsname\relax
  \def\bibfnamefont#1{#1}\fi
\expandafter\ifx\csname citenamefont\endcsname\relax
  \def\citenamefont#1{#1}\fi
\expandafter\ifx\csname url\endcsname\relax
  \def\url#1{\texttt{#1}}\fi
\expandafter\ifx\csname urlprefix\endcsname\relax\def\urlprefix{URL }\fi
\providecommand{\bibinfo}[2]{#2}
\providecommand{\eprint}[2][]{\url{#2}}

\bibitem[{\citenamefont{Abe et~al.}(2015{\natexlab{a}})}]{Abe:2015zbg}
\bibinfo{author}{\bibfnamefont{K.}~\bibnamefont{Abe}} \bibnamefont{et~al.}
  (\bibinfo{collaboration}{Hyper-Kamiokande Proto-Collaboration}), ``{Physics
  potential of a long-baseline neutrino oscillation experiment using a J-PARC
  neutrino beam and Hyper-Kamiokande},'' \bibinfo{journal}{PTEP}
  \textbf{\bibinfo{volume}{2015}}, \bibinfo{pages}{053C02}
  (\bibinfo{year}{2015}{\natexlab{a}}), \eprint{1502.05199}.

\bibitem[{\citenamefont{Abe et~al.}(2016)}]{Abe:2016ero}
\bibinfo{author}{\bibfnamefont{K.}~\bibnamefont{Abe}} \bibnamefont{et~al.}
  (\bibinfo{collaboration}{Hyper-Kamiokande proto-}), ``{Physics Potentials
  with the Second Hyper-Kamiokande Detector in Korea},''
  (\bibinfo{year}{2016}), \eprint{1611.06118}.

\bibitem[{\citenamefont{Adams et~al.}(2013)}]{Adams:2013qkq}
\bibinfo{author}{\bibfnamefont{C.}~\bibnamefont{Adams}} \bibnamefont{et~al.}
  (\bibinfo{collaboration}{LBNE}), ``{The Long-Baseline Neutrino Experiment:
  Exploring Fundamental Symmetries of the Universe},''  (\bibinfo{year}{2013}),
  \eprint{1307.7335}.

\bibitem[{\citenamefont{Acciarri et~al.}(2015)}]{Acciarri:2015uup}
\bibinfo{author}{\bibfnamefont{R.}~\bibnamefont{Acciarri}} \bibnamefont{et~al.}
  (\bibinfo{collaboration}{DUNE}), ``{Long-Baseline Neutrino Facility (LBNF)
  and Deep Underground Neutrino Experiment (DUNE)},''  (\bibinfo{year}{2015}),
  \eprint{1512.06148}.

\bibitem[{\citenamefont{Antusch et~al.}(2006)\citenamefont{Antusch, Biggio,
  Fernandez-Martinez, Gavela, and Lopez-Pavon}}]{Antusch:2006vwa}
\bibinfo{author}{\bibfnamefont{S.}~\bibnamefont{Antusch}},
  \bibinfo{author}{\bibfnamefont{C.}~\bibnamefont{Biggio}},
  \bibinfo{author}{\bibfnamefont{E.}~\bibnamefont{Fernandez-Martinez}},
  \bibinfo{author}{\bibfnamefont{M.~B.} \bibnamefont{Gavela}},
  \bibnamefont{and}
  \bibinfo{author}{\bibfnamefont{J.}~\bibnamefont{Lopez-Pavon}}, ``{Unitarity
  of the Leptonic Mixing Matrix},'' \bibinfo{journal}{JHEP}
  \textbf{\bibinfo{volume}{10}}, \bibinfo{pages}{084} (\bibinfo{year}{2006}),
  \eprint{hep-ph/0607020}.

\bibitem[{\citenamefont{Qian et~al.}(2013)\citenamefont{Qian, Zhang, Diwan, and
  Vogel}}]{Qian:2013ora}
\bibinfo{author}{\bibfnamefont{X.}~\bibnamefont{Qian}},
  \bibinfo{author}{\bibfnamefont{C.}~\bibnamefont{Zhang}},
  \bibinfo{author}{\bibfnamefont{M.}~\bibnamefont{Diwan}}, \bibnamefont{and}
  \bibinfo{author}{\bibfnamefont{P.}~\bibnamefont{Vogel}}, ``{Unitarity Tests
  of the Neutrino Mixing Matrix},''  (\bibinfo{year}{2013}),
  \eprint{1308.5700}.

\bibitem[{\citenamefont{Parke and Ross-Lonergan}(2016)}]{Parke:2015goa}
\bibinfo{author}{\bibfnamefont{S.}~\bibnamefont{Parke}} \bibnamefont{and}
  \bibinfo{author}{\bibfnamefont{M.}~\bibnamefont{Ross-Lonergan}}, ``{Unitarity
  and the three flavor neutrino mixing matrix},'' \bibinfo{journal}{Phys. Rev.}
  \textbf{\bibinfo{volume}{D93}}, \bibinfo{pages}{113009}
  (\bibinfo{year}{2016}), \eprint{1508.05095}.

\bibitem[{\citenamefont{Caldwell and Mohapatra}(1993)}]{Caldwell:1993kn}
\bibinfo{author}{\bibfnamefont{D.~O.} \bibnamefont{Caldwell}} \bibnamefont{and}
  \bibinfo{author}{\bibfnamefont{R.~N.} \bibnamefont{Mohapatra}}, ``{Neutrino
  mass explanations of solar and atmospheric neutrino deficits and hot dark
  matter},'' \bibinfo{journal}{Phys. Rev.} \textbf{\bibinfo{volume}{D48}},
  \bibinfo{pages}{3259} (\bibinfo{year}{1993}).

\bibitem[{\citenamefont{Aguilar-Arevalo et~al.}(2001)}]{Aguilar:2001ty}
\bibinfo{author}{\bibfnamefont{A.}~\bibnamefont{Aguilar-Arevalo}}
  \bibnamefont{et~al.} (\bibinfo{collaboration}{LSND}), ``{Evidence for
  neutrino oscillations from the observation of anti-neutrino(electron)
  appearance in a anti-neutrino(muon) beam},'' \bibinfo{journal}{Phys. Rev.}
  \textbf{\bibinfo{volume}{D64}}, \bibinfo{pages}{112007}
  (\bibinfo{year}{2001}), \eprint{hep-ex/0104049}.

\bibitem[{\citenamefont{Aguilar-Arevalo et~al.}(2009)}]{AguilarArevalo:2008rc}
\bibinfo{author}{\bibfnamefont{A.~A.} \bibnamefont{Aguilar-Arevalo}}
  \bibnamefont{et~al.} (\bibinfo{collaboration}{MiniBooNE}), ``{Unexplained
  Excess of Electron-Like Events From a 1-GeV Neutrino Beam},''
  \bibinfo{journal}{Phys. Rev. Lett.} \textbf{\bibinfo{volume}{102}},
  \bibinfo{pages}{101802} (\bibinfo{year}{2009}), \eprint{0812.2243}.

\bibitem[{\citenamefont{Mention et~al.}(2011)\citenamefont{Mention, Fechner,
  Lasserre, Mueller, Lhuillier, Cribier, and Letourneau}}]{Mention:2011rk}
\bibinfo{author}{\bibfnamefont{G.}~\bibnamefont{Mention}},
  \bibinfo{author}{\bibfnamefont{M.}~\bibnamefont{Fechner}},
  \bibinfo{author}{\bibfnamefont{T.}~\bibnamefont{Lasserre}},
  \bibinfo{author}{\bibfnamefont{T.~A.} \bibnamefont{Mueller}},
  \bibinfo{author}{\bibfnamefont{D.}~\bibnamefont{Lhuillier}},
  \bibinfo{author}{\bibfnamefont{M.}~\bibnamefont{Cribier}}, \bibnamefont{and}
  \bibinfo{author}{\bibfnamefont{A.}~\bibnamefont{Letourneau}}, ``{The Reactor
  Antineutrino Anomaly},'' \bibinfo{journal}{Phys. Rev.}
  \textbf{\bibinfo{volume}{D83}}, \bibinfo{pages}{073006}
  (\bibinfo{year}{2011}), \eprint{1101.2755}.

\bibitem[{\citenamefont{Frekers et~al.}(2011)}]{Frekers:2011zz}
\bibinfo{author}{\bibfnamefont{D.}~\bibnamefont{Frekers}} \bibnamefont{et~al.},
  ``{The Ga-71(He-3, t) reaction and the low-energy neutrino response},''
  \bibinfo{journal}{Phys. Lett.} \textbf{\bibinfo{volume}{B706}},
  \bibinfo{pages}{134} (\bibinfo{year}{2011}).

\bibitem[{\citenamefont{Aguilar-Arevalo
  et~al.}(2012)}]{Aguilar-Arevalo:2012fmn}
\bibinfo{author}{\bibfnamefont{A.~A.} \bibnamefont{Aguilar-Arevalo}}
  \bibnamefont{et~al.} (\bibinfo{collaboration}{MiniBooNE})
  (\bibinfo{year}{2012}), \eprint{1207.4809},
  \urlprefix\url{http://lss.fnal.gov/archive/2012/pub/fermilab-pub-12-394-ad-ppd.pdf}.

\bibitem[{\citenamefont{Aguilar-Arevalo
  et~al.}(2013)}]{Aguilar-Arevalo:2013pmq}
\bibinfo{author}{\bibfnamefont{A.~A.} \bibnamefont{Aguilar-Arevalo}}
  \bibnamefont{et~al.} (\bibinfo{collaboration}{MiniBooNE}), ``{Improved Search
  for $\bar \nu_\mu \rightarrow \bar \nu_e$ Oscillations in the MiniBooNE
  Experiment},'' \bibinfo{journal}{Phys. Rev. Lett.}
  \textbf{\bibinfo{volume}{110}}, \bibinfo{pages}{161801}
  (\bibinfo{year}{2013}), \eprint{1303.2588}.

\bibitem[{\citenamefont{de~Gouv\^ea et~al.}(2015)\citenamefont{de~Gouv\^ea,
  Kelly, and Kobach}}]{deGouvea:2014aoa}
\bibinfo{author}{\bibfnamefont{A.}~\bibnamefont{de~Gouv\^ea}},
  \bibinfo{author}{\bibfnamefont{K.~J.} \bibnamefont{Kelly}}, \bibnamefont{and}
  \bibinfo{author}{\bibfnamefont{A.}~\bibnamefont{Kobach}}, ``{$CP$-invariance
  violation at short-baseline experiments in 3$+$1 neutrino scenarios},''
  \bibinfo{journal}{Phys. Rev.} \textbf{\bibinfo{volume}{D91}},
  \bibinfo{pages}{053005} (\bibinfo{year}{2015}), \eprint{1412.1479}.

\bibitem[{\citenamefont{Adey et~al.}(2015)\citenamefont{Adey, Bayes, Bross, and
  Snopok}}]{Adey:2015iha}
\bibinfo{author}{\bibfnamefont{D.}~\bibnamefont{Adey}},
  \bibinfo{author}{\bibfnamefont{R.}~\bibnamefont{Bayes}},
  \bibinfo{author}{\bibfnamefont{A.}~\bibnamefont{Bross}}, \bibnamefont{and}
  \bibinfo{author}{\bibfnamefont{P.}~\bibnamefont{Snopok}}, ``{nuSTORM and A
  Path to a Muon Collider},'' \bibinfo{journal}{Ann. Rev. Nucl. Part. Sci.}
  \textbf{\bibinfo{volume}{65}}, \bibinfo{pages}{145} (\bibinfo{year}{2015}).

\bibitem[{\citenamefont{Gariazzo et~al.}(2016)\citenamefont{Gariazzo, Giunti,
  Laveder, Li, and Zavanin}}]{Gariazzo:2015rra}
\bibinfo{author}{\bibfnamefont{S.}~\bibnamefont{Gariazzo}},
  \bibinfo{author}{\bibfnamefont{C.}~\bibnamefont{Giunti}},
  \bibinfo{author}{\bibfnamefont{M.}~\bibnamefont{Laveder}},
  \bibinfo{author}{\bibfnamefont{Y.~F.} \bibnamefont{Li}}, \bibnamefont{and}
  \bibinfo{author}{\bibfnamefont{E.~M.} \bibnamefont{Zavanin}}, ``{Light
  sterile neutrinos},'' \bibinfo{journal}{J. Phys.}
  \textbf{\bibinfo{volume}{G43}}, \bibinfo{pages}{033001}
  (\bibinfo{year}{2016}), \eprint{1507.08204}.

\bibitem[{\citenamefont{Giunti}(2016)}]{Giunti:2015wnd}
\bibinfo{author}{\bibfnamefont{C.}~\bibnamefont{Giunti}}, ``{Light Sterile
  Neutrinos: Status and Perspectives},'' \bibinfo{journal}{Nucl. Phys.}
  \textbf{\bibinfo{volume}{B908}}, \bibinfo{pages}{336} (\bibinfo{year}{2016}),
  \eprint{1512.04758}.

\bibitem[{\citenamefont{Choubey and Pramanik}(2017)}]{Choubey:2016fpi}
\bibinfo{author}{\bibfnamefont{S.}~\bibnamefont{Choubey}} \bibnamefont{and}
  \bibinfo{author}{\bibfnamefont{D.}~\bibnamefont{Pramanik}}, ``{Constraints on
  Sterile Neutrino Oscillations using DUNE Near Detector},''
  \bibinfo{journal}{Phys. Lett.} \textbf{\bibinfo{volume}{B764}},
  \bibinfo{pages}{135} (\bibinfo{year}{2017}), \eprint{1604.04731}.

\bibitem[{\citenamefont{Donini et~al.}(2007)\citenamefont{Donini, Maltoni,
  Meloni, Migliozzi, and Terranova}}]{Donini:2007yf}
\bibinfo{author}{\bibfnamefont{A.}~\bibnamefont{Donini}},
  \bibinfo{author}{\bibfnamefont{M.}~\bibnamefont{Maltoni}},
  \bibinfo{author}{\bibfnamefont{D.}~\bibnamefont{Meloni}},
  \bibinfo{author}{\bibfnamefont{P.}~\bibnamefont{Migliozzi}},
  \bibnamefont{and}
  \bibinfo{author}{\bibfnamefont{F.}~\bibnamefont{Terranova}}, ``{3+1 sterile
  neutrinos at the CNGS},'' \bibinfo{journal}{JHEP}
  \textbf{\bibinfo{volume}{12}}, \bibinfo{pages}{013} (\bibinfo{year}{2007}),
  \eprint{0704.0388}.

\bibitem[{\citenamefont{Dighe and Ray}(2007)}]{Dighe:2007uf}
\bibinfo{author}{\bibfnamefont{A.}~\bibnamefont{Dighe}} \bibnamefont{and}
  \bibinfo{author}{\bibfnamefont{S.}~\bibnamefont{Ray}}, ``{Signatures of heavy
  sterile neutrinos at long baseline experiments},'' \bibinfo{journal}{Phys.
  Rev.} \textbf{\bibinfo{volume}{D76}}, \bibinfo{pages}{113001}
  (\bibinfo{year}{2007}), \eprint{0709.0383}.

\bibitem[{\citenamefont{de~Gouv\^ea and Wytock}(2009)}]{deGouvea:2008qk}
\bibinfo{author}{\bibfnamefont{A.}~\bibnamefont{de~Gouv\^ea}} \bibnamefont{and}
  \bibinfo{author}{\bibfnamefont{T.}~\bibnamefont{Wytock}}, ``{Light Sterile
  Neutrino Effects at theta(3)-Sensitive Reactor Neutrino Experiments},''
  \bibinfo{journal}{Phys. Rev.} \textbf{\bibinfo{volume}{D79}},
  \bibinfo{pages}{073005} (\bibinfo{year}{2009}), \eprint{0809.5076}.

\bibitem[{\citenamefont{Meloni et~al.}(2010)\citenamefont{Meloni, Tang, and
  Winter}}]{Meloni:2010zr}
\bibinfo{author}{\bibfnamefont{D.}~\bibnamefont{Meloni}},
  \bibinfo{author}{\bibfnamefont{J.}~\bibnamefont{Tang}}, \bibnamefont{and}
  \bibinfo{author}{\bibfnamefont{W.}~\bibnamefont{Winter}}, ``{Sterile
  neutrinos beyond LSND at the Neutrino Factory},'' \bibinfo{journal}{Phys.
  Rev.} \textbf{\bibinfo{volume}{D82}}, \bibinfo{pages}{093008}
  (\bibinfo{year}{2010}), \eprint{1007.2419}.

\bibitem[{\citenamefont{Bhattacharya et~al.}(2012)\citenamefont{Bhattacharya,
  Thalapillil, and Wagner}}]{Bhattacharya:2011ee}
\bibinfo{author}{\bibfnamefont{B.}~\bibnamefont{Bhattacharya}},
  \bibinfo{author}{\bibfnamefont{A.~M.} \bibnamefont{Thalapillil}},
  \bibnamefont{and} \bibinfo{author}{\bibfnamefont{C.~E.~M.}
  \bibnamefont{Wagner}}, ``{Implications of sterile neutrinos for
  medium/long-baseline neutrino experiments and the determination of
  $\theta_{13}$},'' \bibinfo{journal}{Phys. Rev.}
  \textbf{\bibinfo{volume}{D85}}, \bibinfo{pages}{073004}
  (\bibinfo{year}{2012}), \eprint{1111.4225}.

\bibitem[{\citenamefont{Hollander and Mocioiu}(2015)}]{Hollander:2014iha}
\bibinfo{author}{\bibfnamefont{D.}~\bibnamefont{Hollander}} \bibnamefont{and}
  \bibinfo{author}{\bibfnamefont{I.}~\bibnamefont{Mocioiu}}, ``{Minimal 3+2
  sterile neutrino model at LBNE},'' \bibinfo{journal}{Phys. Rev.}
  \textbf{\bibinfo{volume}{D91}}, \bibinfo{pages}{013002}
  (\bibinfo{year}{2015}), \eprint{1408.1749}.

\bibitem[{\citenamefont{Berryman et~al.}(2015)\citenamefont{Berryman,
  de~Gouvêa, Kelly, and Kobach}}]{Berryman:2015nua}
\bibinfo{author}{\bibfnamefont{J.~M.} \bibnamefont{Berryman}},
  \bibinfo{author}{\bibfnamefont{A.}~\bibnamefont{de~Gouvêa}},
  \bibinfo{author}{\bibfnamefont{K.~J.} \bibnamefont{Kelly}}, \bibnamefont{and}
  \bibinfo{author}{\bibfnamefont{A.}~\bibnamefont{Kobach}}, ``{Sterile neutrino
  at the Deep Underground Neutrino Experiment},'' \bibinfo{journal}{Phys. Rev.}
  \textbf{\bibinfo{volume}{D92}}, \bibinfo{pages}{073012}
  (\bibinfo{year}{2015}), \eprint{1507.03986}.

\bibitem[{\citenamefont{Tabrizi and Peres}(2016)}]{Tabrizi:2015bba}
\bibinfo{author}{\bibfnamefont{Z.}~\bibnamefont{Tabrizi}} \bibnamefont{and}
  \bibinfo{author}{\bibfnamefont{O.~L.~G.} \bibnamefont{Peres}}, ``{Hidden
  Interactions of Sterile Neutrinos As a Probe For New Physics},''
  \bibinfo{journal}{Phys. Rev.} \textbf{\bibinfo{volume}{D93}},
  \bibinfo{pages}{053003} (\bibinfo{year}{2016}), \eprint{1507.06486}.

\bibitem[{\citenamefont{Gandhi et~al.}(2015)\citenamefont{Gandhi, Kayser,
  Masud, and Prakash}}]{Gandhi:2015xza}
\bibinfo{author}{\bibfnamefont{R.}~\bibnamefont{Gandhi}},
  \bibinfo{author}{\bibfnamefont{B.}~\bibnamefont{Kayser}},
  \bibinfo{author}{\bibfnamefont{M.}~\bibnamefont{Masud}}, \bibnamefont{and}
  \bibinfo{author}{\bibfnamefont{S.}~\bibnamefont{Prakash}}, ``{The impact of
  sterile neutrinos on CP measurements at long baselines},''
  \bibinfo{journal}{JHEP} \textbf{\bibinfo{volume}{11}}, \bibinfo{pages}{039}
  (\bibinfo{year}{2015}), \eprint{1508.06275}.

\bibitem[{\citenamefont{Palazzo}(2016)}]{Palazzo:2015gja}
\bibinfo{author}{\bibfnamefont{A.}~\bibnamefont{Palazzo}}, ``{3-flavor and
  4-flavor implications of the latest T2K and NO$\nu$A electron (anti-)neutrino
  appearance results},'' \bibinfo{journal}{Phys. Lett.}
  \textbf{\bibinfo{volume}{B757}}, \bibinfo{pages}{142} (\bibinfo{year}{2016}),
  \eprint{1509.03148}.

\bibitem[{\citenamefont{de~Gouv\^ea and Kobach}(2016)}]{deGouvea:2015euy}
\bibinfo{author}{\bibfnamefont{A.}~\bibnamefont{de~Gouv\^ea}} \bibnamefont{and}
  \bibinfo{author}{\bibfnamefont{A.}~\bibnamefont{Kobach}}, ``{Global
  Constraints on a Heavy Neutrino},'' \bibinfo{journal}{Phys. Rev.}
  \textbf{\bibinfo{volume}{D93}}, \bibinfo{pages}{033005}
  (\bibinfo{year}{2016}), \eprint{1511.00683}.

\bibitem[{\citenamefont{Agarwalla
  et~al.}(2016{\natexlab{a}})\citenamefont{Agarwalla, Chatterjee, Dasgupta, and
  Palazzo}}]{Agarwalla:2016mrc}
\bibinfo{author}{\bibfnamefont{S.~K.} \bibnamefont{Agarwalla}},
  \bibinfo{author}{\bibfnamefont{S.~S.} \bibnamefont{Chatterjee}},
  \bibinfo{author}{\bibfnamefont{A.}~\bibnamefont{Dasgupta}}, \bibnamefont{and}
  \bibinfo{author}{\bibfnamefont{A.}~\bibnamefont{Palazzo}}, ``{Discovery
  Potential of T2K and NOvA in the Presence of a Light Sterile Neutrino},''
  \bibinfo{journal}{JHEP} \textbf{\bibinfo{volume}{02}}, \bibinfo{pages}{111}
  (\bibinfo{year}{2016}{\natexlab{a}}), \eprint{1601.05995}.

\bibitem[{\citenamefont{Agarwalla
  et~al.}(2016{\natexlab{b}})\citenamefont{Agarwalla, Chatterjee, and
  Palazzo}}]{Agarwalla:2016xxa}
\bibinfo{author}{\bibfnamefont{S.~K.} \bibnamefont{Agarwalla}},
  \bibinfo{author}{\bibfnamefont{S.~S.} \bibnamefont{Chatterjee}},
  \bibnamefont{and} \bibinfo{author}{\bibfnamefont{A.}~\bibnamefont{Palazzo}},
  ``{Physics Reach of DUNE with a Light Sterile Neutrino},''
  \bibinfo{journal}{JHEP} \textbf{\bibinfo{volume}{09}}, \bibinfo{pages}{016}
  (\bibinfo{year}{2016}{\natexlab{b}}), \eprint{1603.03759}.

\bibitem[{\citenamefont{Dutta et~al.}(2016)\citenamefont{Dutta, Gandhi, Kayser,
  Masud, and Prakash}}]{Dutta:2016glq}
\bibinfo{author}{\bibfnamefont{D.}~\bibnamefont{Dutta}},
  \bibinfo{author}{\bibfnamefont{R.}~\bibnamefont{Gandhi}},
  \bibinfo{author}{\bibfnamefont{B.}~\bibnamefont{Kayser}},
  \bibinfo{author}{\bibfnamefont{M.}~\bibnamefont{Masud}}, \bibnamefont{and}
  \bibinfo{author}{\bibfnamefont{S.}~\bibnamefont{Prakash}}, ``{Capabilities of
  long-baseline experiments in the presence of a sterile neutrino},''
  \bibinfo{journal}{JHEP} \textbf{\bibinfo{volume}{11}}, \bibinfo{pages}{122}
  (\bibinfo{year}{2016}), \eprint{1607.02152}.

\bibitem[{\citenamefont{Blennow
  et~al.}(2016{\natexlab{a}})\citenamefont{Blennow, Coloma, Fernandez-Martinez,
  Hernandez-Garcia, and Lopez-Pavon}}]{Blennow:2016jkn}
\bibinfo{author}{\bibfnamefont{M.}~\bibnamefont{Blennow}},
  \bibinfo{author}{\bibfnamefont{P.}~\bibnamefont{Coloma}},
  \bibinfo{author}{\bibfnamefont{E.}~\bibnamefont{Fernandez-Martinez}},
  \bibinfo{author}{\bibfnamefont{J.}~\bibnamefont{Hernandez-Garcia}},
  \bibnamefont{and}
  \bibinfo{author}{\bibfnamefont{J.}~\bibnamefont{Lopez-Pavon}},
  ``{Non-Unitarity, sterile neutrinos, and Non-Standard neutrino
  Interactions},''  (\bibinfo{year}{2016}{\natexlab{a}}), \eprint{1609.08637}.

\bibitem[{\citenamefont{Atre et~al.}(2009)\citenamefont{Atre, Han, Pascoli, and
  Zhang}}]{Atre:2009rg}
\bibinfo{author}{\bibfnamefont{A.}~\bibnamefont{Atre}},
  \bibinfo{author}{\bibfnamefont{T.}~\bibnamefont{Han}},
  \bibinfo{author}{\bibfnamefont{S.}~\bibnamefont{Pascoli}}, \bibnamefont{and}
  \bibinfo{author}{\bibfnamefont{B.}~\bibnamefont{Zhang}}, ``{The Search for
  Heavy Majorana Neutrinos},'' \bibinfo{journal}{JHEP}
  \textbf{\bibinfo{volume}{05}}, \bibinfo{pages}{030} (\bibinfo{year}{2009}),
  \eprint{0901.3589}.

\bibitem[{\citenamefont{Vincent et~al.}(2015)\citenamefont{Vincent, Martinez,
  Hernández, Lattanzi, and Mena}}]{Vincent:2014rja}
\bibinfo{author}{\bibfnamefont{A.~C.} \bibnamefont{Vincent}},
  \bibinfo{author}{\bibfnamefont{E.~F.} \bibnamefont{Martinez}},
  \bibinfo{author}{\bibfnamefont{P.}~\bibnamefont{Hernández}},
  \bibinfo{author}{\bibfnamefont{M.}~\bibnamefont{Lattanzi}}, \bibnamefont{and}
  \bibinfo{author}{\bibfnamefont{O.}~\bibnamefont{Mena}}, ``{Revisiting
  cosmological bounds on sterile neutrinos},'' \bibinfo{journal}{JCAP}
  \textbf{\bibinfo{volume}{1504}}, \bibinfo{pages}{006} (\bibinfo{year}{2015}),
  \eprint{1408.1956}.

\bibitem[{\citenamefont{Drewes and Garbrecht}(2015)}]{Drewes:2015iva}
\bibinfo{author}{\bibfnamefont{M.}~\bibnamefont{Drewes}} \bibnamefont{and}
  \bibinfo{author}{\bibfnamefont{B.}~\bibnamefont{Garbrecht}}, ``{Experimental
  and cosmological constraints on heavy neutrinos},''  (\bibinfo{year}{2015}),
  \eprint{1502.00477}.

\bibitem[{\citenamefont{Deppisch et~al.}(2015)\citenamefont{Deppisch,
  Bhupal~Dev, and Pilaftsis}}]{Deppisch:2015qwa}
\bibinfo{author}{\bibfnamefont{F.~F.} \bibnamefont{Deppisch}},
  \bibinfo{author}{\bibfnamefont{P.~S.} \bibnamefont{Bhupal~Dev}},
  \bibnamefont{and}
  \bibinfo{author}{\bibfnamefont{A.}~\bibnamefont{Pilaftsis}}, ``{Neutrinos and
  Collider Physics},'' \bibinfo{journal}{New J. Phys.}
  \textbf{\bibinfo{volume}{17}}, \bibinfo{pages}{075019}
  (\bibinfo{year}{2015}), \eprint{1502.06541}.

\bibitem[{\citenamefont{Adhikari et~al.}(2017)}]{Adhikari:2016bei}
\bibinfo{author}{\bibfnamefont{R.}~\bibnamefont{Adhikari}}
  \bibnamefont{et~al.}, ``{A White Paper on keV Sterile Neutrino Dark
  Matter},'' \bibinfo{journal}{JCAP} \textbf{\bibinfo{volume}{1701}},
  \bibinfo{pages}{025} (\bibinfo{year}{2017}), \eprint{1602.04816}.

\bibitem[{\citenamefont{Wolfenstein}(1978)}]{Wolfenstein:1977ue}
\bibinfo{author}{\bibfnamefont{L.}~\bibnamefont{Wolfenstein}}, ``{Neutrino
  Oscillations in Matter},'' \bibinfo{journal}{Phys. Rev.}
  \textbf{\bibinfo{volume}{D17}}, \bibinfo{pages}{2369} (\bibinfo{year}{1978}).

\bibitem[{\citenamefont{Guzzo et~al.}(1991)\citenamefont{Guzzo, Masiero, and
  Petcov}}]{Guzzo:1991hi}
\bibinfo{author}{\bibfnamefont{M.~M.} \bibnamefont{Guzzo}},
  \bibinfo{author}{\bibfnamefont{A.}~\bibnamefont{Masiero}}, \bibnamefont{and}
  \bibinfo{author}{\bibfnamefont{S.~T.} \bibnamefont{Petcov}}, ``{On the MSW
  effect with massless neutrinos and no mixing in the vacuum},''
  \bibinfo{journal}{Phys. Lett.} \textbf{\bibinfo{volume}{B260}},
  \bibinfo{pages}{154} (\bibinfo{year}{1991}).

\bibitem[{\citenamefont{Krastev and Petcov}(1993)}]{Krastev:1992zx}
\bibinfo{author}{\bibfnamefont{P.~I.} \bibnamefont{Krastev}} \bibnamefont{and}
  \bibinfo{author}{\bibfnamefont{S.~T.} \bibnamefont{Petcov}}, ``{Recent solar
  neutrino observations and unconventional neutrino properties},''
  \bibinfo{journal}{Phys. Lett.} \textbf{\bibinfo{volume}{B299}},
  \bibinfo{pages}{99} (\bibinfo{year}{1993}).

\bibitem[{\citenamefont{Friedland
  et~al.}(2004{\natexlab{a}})\citenamefont{Friedland, Lunardini, and
  Pena-Garay}}]{Friedland:2004pp}
\bibinfo{author}{\bibfnamefont{A.}~\bibnamefont{Friedland}},
  \bibinfo{author}{\bibfnamefont{C.}~\bibnamefont{Lunardini}},
  \bibnamefont{and}
  \bibinfo{author}{\bibfnamefont{C.}~\bibnamefont{Pena-Garay}}, ``{Solar
  neutrinos as probes of neutrino matter interactions},''
  \bibinfo{journal}{Phys. Lett.} \textbf{\bibinfo{volume}{B594}},
  \bibinfo{pages}{347} (\bibinfo{year}{2004}{\natexlab{a}}),
  \eprint{hep-ph/0402266}.

\bibitem[{\citenamefont{Miranda et~al.}(2006)\citenamefont{Miranda, Tortola,
  and Valle}}]{Miranda:2004nb}
\bibinfo{author}{\bibfnamefont{O.~G.} \bibnamefont{Miranda}},
  \bibinfo{author}{\bibfnamefont{M.~A.} \bibnamefont{Tortola}},
  \bibnamefont{and} \bibinfo{author}{\bibfnamefont{J.~W.~F.}
  \bibnamefont{Valle}}, ``{Are solar neutrino oscillations robust?},''
  \bibinfo{journal}{JHEP} \textbf{\bibinfo{volume}{10}}, \bibinfo{pages}{008}
  (\bibinfo{year}{2006}), \eprint{hep-ph/0406280}.

\bibitem[{\citenamefont{Bolanos et~al.}(2009)\citenamefont{Bolanos, Miranda,
  Palazzo, Tortola, and Valle}}]{Bolanos:2008km}
\bibinfo{author}{\bibfnamefont{A.}~\bibnamefont{Bolanos}},
  \bibinfo{author}{\bibfnamefont{O.~G.} \bibnamefont{Miranda}},
  \bibinfo{author}{\bibfnamefont{A.}~\bibnamefont{Palazzo}},
  \bibinfo{author}{\bibfnamefont{M.~A.} \bibnamefont{Tortola}},
  \bibnamefont{and} \bibinfo{author}{\bibfnamefont{J.~W.~F.}
  \bibnamefont{Valle}}, ``{Probing non-standard neutrino-electron interactions
  with solar and reactor neutrinos},'' \bibinfo{journal}{Phys. Rev.}
  \textbf{\bibinfo{volume}{D79}}, \bibinfo{pages}{113012}
  (\bibinfo{year}{2009}), \eprint{0812.4417}.

\bibitem[{\citenamefont{Palazzo and Valle}(2009)}]{Palazzo:2009rb}
\bibinfo{author}{\bibfnamefont{A.}~\bibnamefont{Palazzo}} \bibnamefont{and}
  \bibinfo{author}{\bibfnamefont{J.~W.~F.} \bibnamefont{Valle}}, ``{Confusing
  non-zero $\theta_{13}$ with non-standard interactions in the solar neutrino
  sector},'' \bibinfo{journal}{Phys. Rev.} \textbf{\bibinfo{volume}{D80}},
  \bibinfo{pages}{091301} (\bibinfo{year}{2009}), \eprint{0909.1535}.

\bibitem[{\citenamefont{Escrihuela et~al.}(2009)\citenamefont{Escrihuela,
  Miranda, Tortola, and Valle}}]{Escrihuela:2009up}
\bibinfo{author}{\bibfnamefont{F.~J.} \bibnamefont{Escrihuela}},
  \bibinfo{author}{\bibfnamefont{O.~G.} \bibnamefont{Miranda}},
  \bibinfo{author}{\bibfnamefont{M.~A.} \bibnamefont{Tortola}},
  \bibnamefont{and} \bibinfo{author}{\bibfnamefont{J.~W.~F.}
  \bibnamefont{Valle}}, ``{Constraining nonstandard neutrino-quark interactions
  with solar, reactor and accelerator data},'' \bibinfo{journal}{Phys. Rev.}
  \textbf{\bibinfo{volume}{D80}}, \bibinfo{pages}{105009}
  (\bibinfo{year}{2009}), \bibinfo{note}{[Erratum: Phys.
  Rev.D80,129908(2009)]}, \eprint{0907.2630}.

\bibitem[{\citenamefont{Gonzalez-Garcia
  et~al.}(1999)\citenamefont{Gonzalez-Garcia, Guzzo, Krastev, Nunokawa, Peres,
  Pleitez, Valle, and Zukanovich~Funchal}}]{GonzalezGarcia:1998hj}
\bibinfo{author}{\bibfnamefont{M.~C.} \bibnamefont{Gonzalez-Garcia}},
  \bibinfo{author}{\bibfnamefont{M.~M.} \bibnamefont{Guzzo}},
  \bibinfo{author}{\bibfnamefont{P.~I.} \bibnamefont{Krastev}},
  \bibinfo{author}{\bibfnamefont{H.}~\bibnamefont{Nunokawa}},
  \bibinfo{author}{\bibfnamefont{O.~L.~G.} \bibnamefont{Peres}},
  \bibinfo{author}{\bibfnamefont{V.}~\bibnamefont{Pleitez}},
  \bibinfo{author}{\bibfnamefont{J.~W.~F.} \bibnamefont{Valle}},
  \bibnamefont{and}
  \bibinfo{author}{\bibfnamefont{R.}~\bibnamefont{Zukanovich~Funchal}},
  ``{Atmospheric neutrino observations and flavor changing interactions},''
  \bibinfo{journal}{Phys. Rev. Lett.} \textbf{\bibinfo{volume}{82}},
  \bibinfo{pages}{3202} (\bibinfo{year}{1999}), \eprint{hep-ph/9809531}.

\bibitem[{\citenamefont{Fornengo et~al.}(2000)\citenamefont{Fornengo,
  Gonzalez-Garcia, and Valle}}]{Fornengo:1999zp}
\bibinfo{author}{\bibfnamefont{N.}~\bibnamefont{Fornengo}},
  \bibinfo{author}{\bibfnamefont{M.~C.} \bibnamefont{Gonzalez-Garcia}},
  \bibnamefont{and} \bibinfo{author}{\bibfnamefont{J.~W.~F.}
  \bibnamefont{Valle}}, ``{On the interpretation of the atmospheric neutrino
  data in terms of flavor changing neutrino interactions},''
  \bibinfo{journal}{JHEP} \textbf{\bibinfo{volume}{07}}, \bibinfo{pages}{006}
  (\bibinfo{year}{2000}), \eprint{hep-ph/9906539}.

\bibitem[{\citenamefont{Fornengo et~al.}(2002)\citenamefont{Fornengo, Maltoni,
  Tomas, and Valle}}]{Fornengo:2001pm}
\bibinfo{author}{\bibfnamefont{N.}~\bibnamefont{Fornengo}},
  \bibinfo{author}{\bibfnamefont{M.}~\bibnamefont{Maltoni}},
  \bibinfo{author}{\bibfnamefont{R.}~\bibnamefont{Tomas}}, \bibnamefont{and}
  \bibinfo{author}{\bibfnamefont{J.~W.~F.} \bibnamefont{Valle}}, ``{Probing
  neutrino nonstandard interactions with atmospheric neutrino data},''
  \bibinfo{journal}{Phys. Rev.} \textbf{\bibinfo{volume}{D65}},
  \bibinfo{pages}{013010} (\bibinfo{year}{2002}), \eprint{hep-ph/0108043}.

\bibitem[{\citenamefont{Huber and Valle}(2001)}]{Huber:2001zw}
\bibinfo{author}{\bibfnamefont{P.}~\bibnamefont{Huber}} \bibnamefont{and}
  \bibinfo{author}{\bibfnamefont{J.~W.~F.} \bibnamefont{Valle}}, ``{Nonstandard
  interactions: Atmospheric versus neutrino factory experiments},''
  \bibinfo{journal}{Phys. Lett.} \textbf{\bibinfo{volume}{B523}},
  \bibinfo{pages}{151} (\bibinfo{year}{2001}), \eprint{hep-ph/0108193}.

\bibitem[{\citenamefont{Friedland
  et~al.}(2004{\natexlab{b}})\citenamefont{Friedland, Lunardini, and
  Maltoni}}]{Friedland:2004ah}
\bibinfo{author}{\bibfnamefont{A.}~\bibnamefont{Friedland}},
  \bibinfo{author}{\bibfnamefont{C.}~\bibnamefont{Lunardini}},
  \bibnamefont{and} \bibinfo{author}{\bibfnamefont{M.}~\bibnamefont{Maltoni}},
  ``{Atmospheric neutrinos as probes of neutrino-matter interactions},''
  \bibinfo{journal}{Phys. Rev.} \textbf{\bibinfo{volume}{D70}},
  \bibinfo{pages}{111301} (\bibinfo{year}{2004}{\natexlab{b}}),
  \eprint{hep-ph/0408264}.

\bibitem[{\citenamefont{Friedland and Lunardini}(2005)}]{Friedland:2005vy}
\bibinfo{author}{\bibfnamefont{A.}~\bibnamefont{Friedland}} \bibnamefont{and}
  \bibinfo{author}{\bibfnamefont{C.}~\bibnamefont{Lunardini}}, ``{A Test of tau
  neutrino interactions with atmospheric neutrinos and K2K},''
  \bibinfo{journal}{Phys. Rev.} \textbf{\bibinfo{volume}{D72}},
  \bibinfo{pages}{053009} (\bibinfo{year}{2005}), \eprint{hep-ph/0506143}.

\bibitem[{\citenamefont{Yasuda}(2011)}]{Yasuda:2010hw}
\bibinfo{author}{\bibfnamefont{O.}~\bibnamefont{Yasuda}}, ``{Sensitivity of
  T2KK to non-standard interactions},'' \bibinfo{journal}{Nucl. Phys. Proc.
  Suppl.} \textbf{\bibinfo{volume}{217}}, \bibinfo{pages}{220}
  (\bibinfo{year}{2011}), \eprint{1011.6440}.

\bibitem[{\citenamefont{Gonzalez-Garcia
  et~al.}(2011)\citenamefont{Gonzalez-Garcia, Maltoni, and
  Salvado}}]{GonzalezGarcia:2011my}
\bibinfo{author}{\bibfnamefont{M.~C.} \bibnamefont{Gonzalez-Garcia}},
  \bibinfo{author}{\bibfnamefont{M.}~\bibnamefont{Maltoni}}, \bibnamefont{and}
  \bibinfo{author}{\bibfnamefont{J.}~\bibnamefont{Salvado}}, ``{Testing matter
  effects in propagation of atmospheric and long-baseline neutrinos},''
  \bibinfo{journal}{JHEP} \textbf{\bibinfo{volume}{05}}, \bibinfo{pages}{075}
  (\bibinfo{year}{2011}), \eprint{1103.4365}.

\bibitem[{\citenamefont{Esmaili and Smirnov}(2013)}]{Esmaili:2013fva}
\bibinfo{author}{\bibfnamefont{A.}~\bibnamefont{Esmaili}} \bibnamefont{and}
  \bibinfo{author}{\bibfnamefont{A.~{\relax Yu}.} \bibnamefont{Smirnov}},
  ``{Probing Non-Standard Interaction of Neutrinos with IceCube and
  DeepCore},'' \bibinfo{journal}{JHEP} \textbf{\bibinfo{volume}{06}},
  \bibinfo{pages}{026} (\bibinfo{year}{2013}), \eprint{1304.1042}.

\bibitem[{\citenamefont{Choubey and Ohlsson}(2014)}]{Choubey:2014iia}
\bibinfo{author}{\bibfnamefont{S.}~\bibnamefont{Choubey}} \bibnamefont{and}
  \bibinfo{author}{\bibfnamefont{T.}~\bibnamefont{Ohlsson}}, ``{Bounds on
  Non-Standard Neutrino Interactions Using PINGU},'' \bibinfo{journal}{Phys.
  Lett.} \textbf{\bibinfo{volume}{B739}}, \bibinfo{pages}{357}
  (\bibinfo{year}{2014}), \eprint{1410.0410}.

\bibitem[{\citenamefont{Mocioiu and Wright}(2015)}]{Mocioiu:2014gua}
\bibinfo{author}{\bibfnamefont{I.}~\bibnamefont{Mocioiu}} \bibnamefont{and}
  \bibinfo{author}{\bibfnamefont{W.}~\bibnamefont{Wright}}, ``{Non-standard
  neutrino interactions in the mu–tau sector},'' \bibinfo{journal}{Nucl.
  Phys.} \textbf{\bibinfo{volume}{B893}}, \bibinfo{pages}{376}
  (\bibinfo{year}{2015}), \eprint{1410.6193}.

\bibitem[{\citenamefont{Fukasawa and Yasuda}(2015)}]{Fukasawa:2015jaa}
\bibinfo{author}{\bibfnamefont{S.}~\bibnamefont{Fukasawa}} \bibnamefont{and}
  \bibinfo{author}{\bibfnamefont{O.}~\bibnamefont{Yasuda}}, ``{Constraints on
  the Nonstandard Interaction in Propagation from Atmospheric Neutrinos},''
  \bibinfo{journal}{Adv. High Energy Phys.} \textbf{\bibinfo{volume}{2015}},
  \bibinfo{pages}{820941} (\bibinfo{year}{2015}), \eprint{1503.08056}.

\bibitem[{\citenamefont{Choubey et~al.}(2015)\citenamefont{Choubey, Ghosh,
  Ohlsson, and Tiwari}}]{Choubey:2015xha}
\bibinfo{author}{\bibfnamefont{S.}~\bibnamefont{Choubey}},
  \bibinfo{author}{\bibfnamefont{A.}~\bibnamefont{Ghosh}},
  \bibinfo{author}{\bibfnamefont{T.}~\bibnamefont{Ohlsson}}, \bibnamefont{and}
  \bibinfo{author}{\bibfnamefont{D.}~\bibnamefont{Tiwari}}, ``{Neutrino Physics
  with Non-Standard Interactions at INO},'' \bibinfo{journal}{JHEP}
  \textbf{\bibinfo{volume}{12}}, \bibinfo{pages}{126} (\bibinfo{year}{2015}),
  \eprint{1507.02211}.

\bibitem[{\citenamefont{Salvado et~al.}(2017)\citenamefont{Salvado, Mena,
  Palomares-Ruiz, and Rius}}]{Salvado:2016uqu}
\bibinfo{author}{\bibfnamefont{J.}~\bibnamefont{Salvado}},
  \bibinfo{author}{\bibfnamefont{O.}~\bibnamefont{Mena}},
  \bibinfo{author}{\bibfnamefont{S.}~\bibnamefont{Palomares-Ruiz}},
  \bibnamefont{and} \bibinfo{author}{\bibfnamefont{N.}~\bibnamefont{Rius}},
  ``{Non-standard interactions with high-energy atmospheric neutrinos at
  IceCube},'' \bibinfo{journal}{JHEP} \textbf{\bibinfo{volume}{01}},
  \bibinfo{pages}{141} (\bibinfo{year}{2017}), \eprint{1609.03450}.

\bibitem[{\citenamefont{Friedland and Lunardini}(2006)}]{Friedland:2006pi}
\bibinfo{author}{\bibfnamefont{A.}~\bibnamefont{Friedland}} \bibnamefont{and}
  \bibinfo{author}{\bibfnamefont{C.}~\bibnamefont{Lunardini}}, ``{Two modes of
  searching for new neutrino interactions at MINOS},'' \bibinfo{journal}{Phys.
  Rev.} \textbf{\bibinfo{volume}{D74}}, \bibinfo{pages}{033012}
  (\bibinfo{year}{2006}), \eprint{hep-ph/0606101}.

\bibitem[{\citenamefont{Blennow et~al.}(2008)\citenamefont{Blennow, Ohlsson,
  and Skrotzki}}]{Blennow:2007pu}
\bibinfo{author}{\bibfnamefont{M.}~\bibnamefont{Blennow}},
  \bibinfo{author}{\bibfnamefont{T.}~\bibnamefont{Ohlsson}}, \bibnamefont{and}
  \bibinfo{author}{\bibfnamefont{J.}~\bibnamefont{Skrotzki}}, ``{Effects of
  non-standard interactions in the MINOS experiment},'' \bibinfo{journal}{Phys.
  Lett.} \textbf{\bibinfo{volume}{B660}}, \bibinfo{pages}{522}
  (\bibinfo{year}{2008}), \eprint{hep-ph/0702059}.

\bibitem[{\citenamefont{Esteban-Pretel
  et~al.}(2008)\citenamefont{Esteban-Pretel, Valle, and
  Huber}}]{EstebanPretel:2008qi}
\bibinfo{author}{\bibfnamefont{A.}~\bibnamefont{Esteban-Pretel}},
  \bibinfo{author}{\bibfnamefont{J.~W.~F.} \bibnamefont{Valle}},
  \bibnamefont{and} \bibinfo{author}{\bibfnamefont{P.}~\bibnamefont{Huber}},
  ``{Can OPERA help in constraining neutrino non-standard interactions?},''
  \bibinfo{journal}{Phys. Lett.} \textbf{\bibinfo{volume}{B668}},
  \bibinfo{pages}{197} (\bibinfo{year}{2008}), \eprint{0803.1790}.

\bibitem[{\citenamefont{Kopp et~al.}(2010)\citenamefont{Kopp, Machado, and
  Parke}}]{Kopp:2010qt}
\bibinfo{author}{\bibfnamefont{J.}~\bibnamefont{Kopp}},
  \bibinfo{author}{\bibfnamefont{P.~A.~N.} \bibnamefont{Machado}},
  \bibnamefont{and} \bibinfo{author}{\bibfnamefont{S.~J.} \bibnamefont{Parke}},
  ``{Interpretation of MINOS data in terms of non-standard neutrino
  interactions},'' \bibinfo{journal}{Phys. Rev.}
  \textbf{\bibinfo{volume}{D82}}, \bibinfo{pages}{113002}
  (\bibinfo{year}{2010}), \eprint{1009.0014}.

\bibitem[{\citenamefont{Coloma et~al.}(2011)\citenamefont{Coloma, Donini,
  Lopez-Pavon, and Minakata}}]{Coloma:2011rq}
\bibinfo{author}{\bibfnamefont{P.}~\bibnamefont{Coloma}},
  \bibinfo{author}{\bibfnamefont{A.}~\bibnamefont{Donini}},
  \bibinfo{author}{\bibfnamefont{J.}~\bibnamefont{Lopez-Pavon}},
  \bibnamefont{and} \bibinfo{author}{\bibfnamefont{H.}~\bibnamefont{Minakata}},
  ``{Non-Standard Interactions at a Neutrino Factory: Correlations and CP
  violation},'' \bibinfo{journal}{JHEP} \textbf{\bibinfo{volume}{08}},
  \bibinfo{pages}{036} (\bibinfo{year}{2011}), \eprint{1105.5936}.

\bibitem[{\citenamefont{Friedland and Shoemaker}(2012)}]{Friedland:2012tq}
\bibinfo{author}{\bibfnamefont{A.}~\bibnamefont{Friedland}} \bibnamefont{and}
  \bibinfo{author}{\bibfnamefont{I.~M.} \bibnamefont{Shoemaker}}, ``{Searching
  for Novel Neutrino Interactions at NOvA and Beyond in Light of Large
  $\theta_{13}$},''  (\bibinfo{year}{2012}), \eprint{1207.6642}.

\bibitem[{\citenamefont{Coelho et~al.}(2012)\citenamefont{Coelho, Kafka, Mann,
  Schneps, and Altinok}}]{Coelho:2012bp}
\bibinfo{author}{\bibfnamefont{J.~A.~B.} \bibnamefont{Coelho}},
  \bibinfo{author}{\bibfnamefont{T.}~\bibnamefont{Kafka}},
  \bibinfo{author}{\bibfnamefont{W.~A.} \bibnamefont{Mann}},
  \bibinfo{author}{\bibfnamefont{J.}~\bibnamefont{Schneps}}, \bibnamefont{and}
  \bibinfo{author}{\bibfnamefont{O.}~\bibnamefont{Altinok}}, ``{Constraints for
  non-standard interaction $\epsilon_{e \tau}V_e$ from $\nu_e$ appearance in
  MINOS and T2K},'' \bibinfo{journal}{Phys. Rev.}
  \textbf{\bibinfo{volume}{D86}}, \bibinfo{pages}{113015}
  (\bibinfo{year}{2012}), \eprint{1209.3757}.

\bibitem[{\citenamefont{Adamson et~al.}(2013)}]{Adamson:2013ovz}
\bibinfo{author}{\bibfnamefont{P.}~\bibnamefont{Adamson}} \bibnamefont{et~al.}
  (\bibinfo{collaboration}{MINOS}), ``{Search for flavor-changing non-standard
  neutrino interactions by MINOS},'' \bibinfo{journal}{Phys. Rev.}
  \textbf{\bibinfo{volume}{D88}}, \bibinfo{pages}{072011}
  (\bibinfo{year}{2013}), \eprint{1303.5314}.

\bibitem[{\citenamefont{Girardi et~al.}(2014)\citenamefont{Girardi, Meloni, and
  Petcov}}]{Girardi:2014kca}
\bibinfo{author}{\bibfnamefont{I.}~\bibnamefont{Girardi}},
  \bibinfo{author}{\bibfnamefont{D.}~\bibnamefont{Meloni}}, \bibnamefont{and}
  \bibinfo{author}{\bibfnamefont{S.~T.} \bibnamefont{Petcov}}, ``{The Daya Bay
  and T2K results on $\sin^2 2 \theta_{13}$ and Non-Standard Neutrino
  Interactions},'' \bibinfo{journal}{Nucl. Phys.}
  \textbf{\bibinfo{volume}{B886}}, \bibinfo{pages}{31} (\bibinfo{year}{2014}),
  \eprint{1405.0416}.

\bibitem[{\citenamefont{Blennow et~al.}(2015)\citenamefont{Blennow, Choubey,
  Ohlsson, and Raut}}]{Blennow:2015nxa}
\bibinfo{author}{\bibfnamefont{M.}~\bibnamefont{Blennow}},
  \bibinfo{author}{\bibfnamefont{S.}~\bibnamefont{Choubey}},
  \bibinfo{author}{\bibfnamefont{T.}~\bibnamefont{Ohlsson}}, \bibnamefont{and}
  \bibinfo{author}{\bibfnamefont{S.~K.} \bibnamefont{Raut}}, ``{Exploring
  Source and Detector Non-Standard Neutrino Interactions at ESS$\nu$SB},''
  \bibinfo{journal}{JHEP} \textbf{\bibinfo{volume}{09}}, \bibinfo{pages}{096}
  (\bibinfo{year}{2015}), \eprint{1507.02868}.

\bibitem[{\citenamefont{Masud et~al.}(2016)\citenamefont{Masud, Chatterjee, and
  Mehta}}]{Masud:2015xva}
\bibinfo{author}{\bibfnamefont{M.}~\bibnamefont{Masud}},
  \bibinfo{author}{\bibfnamefont{A.}~\bibnamefont{Chatterjee}},
  \bibnamefont{and} \bibinfo{author}{\bibfnamefont{P.}~\bibnamefont{Mehta}},
  ``{Probing CP violation signal at DUNE in presence of non-standard neutrino
  interactions},'' \bibinfo{journal}{J. Phys.} \textbf{\bibinfo{volume}{G43}},
  \bibinfo{pages}{095005} (\bibinfo{year}{2016}), \eprint{1510.08261}.

\bibitem[{\citenamefont{de~Gouv\^ea and
  Kelly}(2016{\natexlab{a}})}]{deGouvea:2015ndi}
\bibinfo{author}{\bibfnamefont{A.}~\bibnamefont{de~Gouv\^ea}} \bibnamefont{and}
  \bibinfo{author}{\bibfnamefont{K.~J.} \bibnamefont{Kelly}}, ``{Non-standard
  Neutrino Interactions at DUNE},'' \bibinfo{journal}{Nucl. Phys.}
  \textbf{\bibinfo{volume}{B908}}, \bibinfo{pages}{318}
  (\bibinfo{year}{2016}{\natexlab{a}}), \eprint{1511.05562}.

\bibitem[{\citenamefont{Coloma}(2016)}]{Coloma:2015kiu}
\bibinfo{author}{\bibfnamefont{P.}~\bibnamefont{Coloma}}, ``{Non-Standard
  Interactions in propagation at the Deep Underground Neutrino Experiment},''
  \bibinfo{journal}{JHEP} \textbf{\bibinfo{volume}{03}}, \bibinfo{pages}{016}
  (\bibinfo{year}{2016}), \eprint{1511.06357}.

\bibitem[{\citenamefont{Liao et~al.}(2016)\citenamefont{Liao, Marfatia, and
  Whisnant}}]{Liao:2016hsa}
\bibinfo{author}{\bibfnamefont{J.}~\bibnamefont{Liao}},
  \bibinfo{author}{\bibfnamefont{D.}~\bibnamefont{Marfatia}}, \bibnamefont{and}
  \bibinfo{author}{\bibfnamefont{K.}~\bibnamefont{Whisnant}}, ``{Degeneracies
  in long-baseline neutrino experiments from nonstandard interactions},''
  \bibinfo{journal}{Phys. Rev.} \textbf{\bibinfo{volume}{D93}},
  \bibinfo{pages}{093016} (\bibinfo{year}{2016}), \eprint{1601.00927}.

\bibitem[{\citenamefont{Forero and Huber}(2016)}]{Forero:2016cmb}
\bibinfo{author}{\bibfnamefont{D.~V.} \bibnamefont{Forero}} \bibnamefont{and}
  \bibinfo{author}{\bibfnamefont{P.}~\bibnamefont{Huber}}, ``{Hints for
  leptonic CP violation or New Physics?},'' \bibinfo{journal}{Phys. Rev. Lett.}
  \textbf{\bibinfo{volume}{117}}, \bibinfo{pages}{031801}
  (\bibinfo{year}{2016}), \eprint{1601.03736}.

\bibitem[{\citenamefont{Huitu et~al.}(2016)\citenamefont{Huitu,
  K{\"a}rkk{\"a}inen, Maalampi, and Vihonen}}]{Huitu:2016bmb}
\bibinfo{author}{\bibfnamefont{K.}~\bibnamefont{Huitu}},
  \bibinfo{author}{\bibfnamefont{T.~J.} \bibnamefont{K{\"a}rkk{\"a}inen}},
  \bibinfo{author}{\bibfnamefont{J.}~\bibnamefont{Maalampi}}, \bibnamefont{and}
  \bibinfo{author}{\bibfnamefont{S.}~\bibnamefont{Vihonen}}, ``{Constraining
  the nonstandard interaction parameters in long baseline neutrino
  experiments},'' \bibinfo{journal}{Phys. Rev.} \textbf{\bibinfo{volume}{D93}},
  \bibinfo{pages}{053016} (\bibinfo{year}{2016}), \eprint{1601.07730}.

\bibitem[{\citenamefont{Bakhti and Farzan}(2016)}]{Bakhti:2016prn}
\bibinfo{author}{\bibfnamefont{P.}~\bibnamefont{Bakhti}} \bibnamefont{and}
  \bibinfo{author}{\bibfnamefont{Y.}~\bibnamefont{Farzan}}, ``{CP-Violation and
  Non-Standard Interactions at the MOMENT},'' \bibinfo{journal}{JHEP}
  \textbf{\bibinfo{volume}{07}}, \bibinfo{pages}{109} (\bibinfo{year}{2016}),
  \eprint{1602.07099}.

\bibitem[{\citenamefont{Masud and Mehta}(2016{\natexlab{a}})}]{Masud:2016bvp}
\bibinfo{author}{\bibfnamefont{M.}~\bibnamefont{Masud}} \bibnamefont{and}
  \bibinfo{author}{\bibfnamefont{P.}~\bibnamefont{Mehta}}, ``{Nonstandard
  interactions spoiling the CP violation sensitivity at DUNE and other long
  baseline experiments},'' \bibinfo{journal}{Phys. Rev.}
  \textbf{\bibinfo{volume}{D94}}, \bibinfo{pages}{013014}
  (\bibinfo{year}{2016}{\natexlab{a}}), \eprint{1603.01380}.

\bibitem[{\citenamefont{Miranda et~al.}(2016)\citenamefont{Miranda, Tortola,
  and Valle}}]{Miranda:2016wdr}
\bibinfo{author}{\bibfnamefont{O.~G.} \bibnamefont{Miranda}},
  \bibinfo{author}{\bibfnamefont{M.}~\bibnamefont{Tortola}}, \bibnamefont{and}
  \bibinfo{author}{\bibfnamefont{J.~W.~F.} \bibnamefont{Valle}}, ``{New
  ambiguity in probing CP violation in neutrino oscillations},''
  \bibinfo{journal}{Phys. Rev. Lett.} \textbf{\bibinfo{volume}{117}},
  \bibinfo{pages}{061804} (\bibinfo{year}{2016}), \eprint{1604.05690}.

\bibitem[{\citenamefont{Coloma and Schwetz}(2016)}]{Coloma:2016gei}
\bibinfo{author}{\bibfnamefont{P.}~\bibnamefont{Coloma}} \bibnamefont{and}
  \bibinfo{author}{\bibfnamefont{T.}~\bibnamefont{Schwetz}}, ``{Generalized
  mass ordering degeneracy in neutrino oscillation experiments},''
  \bibinfo{journal}{Phys. Rev.} \textbf{\bibinfo{volume}{D94}},
  \bibinfo{pages}{055005} (\bibinfo{year}{2016}), \eprint{1604.05772}.

\bibitem[{\citenamefont{Khan}(2016)}]{Khan:2016uon}
\bibinfo{author}{\bibfnamefont{A.~N.} \bibnamefont{Khan}}, ``{Global analysis
  of the source and detector nonstandard interactions using the short baseline
  $\nu$-e and $\bar{\nu}$-e scattering data},'' \bibinfo{journal}{Phys. Rev.}
  \textbf{\bibinfo{volume}{D93}}, \bibinfo{pages}{093019}
  (\bibinfo{year}{2016}), \eprint{1605.09284}.

\bibitem[{\citenamefont{de~Gouv\^ea and
  Kelly}(2016{\natexlab{b}})}]{deGouvea:2016pom}
\bibinfo{author}{\bibfnamefont{A.}~\bibnamefont{de~Gouv\^ea}} \bibnamefont{and}
  \bibinfo{author}{\bibfnamefont{K.~J.} \bibnamefont{Kelly}}, ``{False Signals
  of CP-Invariance Violation at DUNE},''  (\bibinfo{year}{2016}{\natexlab{b}}),
  \eprint{1605.09376}.

\bibitem[{\citenamefont{Masud and Mehta}(2016{\natexlab{b}})}]{Masud:2016gcl}
\bibinfo{author}{\bibfnamefont{M.}~\bibnamefont{Masud}} \bibnamefont{and}
  \bibinfo{author}{\bibfnamefont{P.}~\bibnamefont{Mehta}}, ``{Nonstandard
  interactions and resolving the ordering of neutrino masses at DUNE and other
  long baseline experiments},'' \bibinfo{journal}{Phys. Rev.}
  \textbf{\bibinfo{volume}{D94}}, \bibinfo{pages}{053007}
  (\bibinfo{year}{2016}{\natexlab{b}}), \eprint{1606.05662}.

\bibitem[{\citenamefont{Blennow
  et~al.}(2016{\natexlab{b}})\citenamefont{Blennow, Choubey, Ohlsson, Pramanik,
  and Raut}}]{Blennow:2016etl}
\bibinfo{author}{\bibfnamefont{M.}~\bibnamefont{Blennow}},
  \bibinfo{author}{\bibfnamefont{S.}~\bibnamefont{Choubey}},
  \bibinfo{author}{\bibfnamefont{T.}~\bibnamefont{Ohlsson}},
  \bibinfo{author}{\bibfnamefont{D.}~\bibnamefont{Pramanik}}, \bibnamefont{and}
  \bibinfo{author}{\bibfnamefont{S.~K.} \bibnamefont{Raut}}, ``{A combined
  study of source, detector and matter non-standard neutrino interactions at
  DUNE},'' \bibinfo{journal}{JHEP} \textbf{\bibinfo{volume}{08}},
  \bibinfo{pages}{090} (\bibinfo{year}{2016}{\natexlab{b}}),
  \eprint{1606.08851}.

\bibitem[{\citenamefont{Bakhti and Khan}(2016)}]{Bakhti:2016gic}
\bibinfo{author}{\bibfnamefont{P.}~\bibnamefont{Bakhti}} \bibnamefont{and}
  \bibinfo{author}{\bibfnamefont{A.~N.} \bibnamefont{Khan}}, ``{Sensitivities
  to charged-current nonstandard neutrino interactions at DUNE},''
  (\bibinfo{year}{2016}), \eprint{1607.00065}.

\bibitem[{\citenamefont{Farzan and Heeck}(2016)}]{Farzan:2016wym}
\bibinfo{author}{\bibfnamefont{Y.}~\bibnamefont{Farzan}} \bibnamefont{and}
  \bibinfo{author}{\bibfnamefont{J.}~\bibnamefont{Heeck}}, ``{Neutrinophilic
  nonstandard interactions},'' \bibinfo{journal}{Phys. Rev.}
  \textbf{\bibinfo{volume}{D94}}, \bibinfo{pages}{053010}
  (\bibinfo{year}{2016}), \eprint{1607.07616}.

\bibitem[{\citenamefont{Forero and Huang}(2016)}]{Forero:2016ghr}
\bibinfo{author}{\bibfnamefont{D.~V.} \bibnamefont{Forero}} \bibnamefont{and}
  \bibinfo{author}{\bibfnamefont{W.-C.} \bibnamefont{Huang}}, ``{Sizable NSI
  from the $SU(2)_L$ scalar doublet-singlet mixing and the implications in
  DUNE},''  (\bibinfo{year}{2016}), \eprint{1608.04719}.

\bibitem[{\citenamefont{Fukasawa
  et~al.}(2016{\natexlab{a}})\citenamefont{Fukasawa, Ghosh, and
  Yasuda}}]{Fukasawa:2016gvm}
\bibinfo{author}{\bibfnamefont{S.}~\bibnamefont{Fukasawa}},
  \bibinfo{author}{\bibfnamefont{M.}~\bibnamefont{Ghosh}}, \bibnamefont{and}
  \bibinfo{author}{\bibfnamefont{O.}~\bibnamefont{Yasuda}}, ``{Is nonstandard
  interaction a solution to the three neutrino tensions?},''
  (\bibinfo{year}{2016}{\natexlab{a}}), \eprint{1609.04204}.

\bibitem[{\citenamefont{Deepthi et~al.}(2016)\citenamefont{Deepthi, Goswami,
  and Nath}}]{Deepthi:2016erc}
\bibinfo{author}{\bibfnamefont{K.~N.} \bibnamefont{Deepthi}},
  \bibinfo{author}{\bibfnamefont{S.}~\bibnamefont{Goswami}}, \bibnamefont{and}
  \bibinfo{author}{\bibfnamefont{N.}~\bibnamefont{Nath}}, ``{Nonstandard
  interactions jeopardizing the hierarchy sensitivity of DUNE},''
  (\bibinfo{year}{2016}), \eprint{1612.00784}.

\bibitem[{\citenamefont{Ge and Smirnov}(2016)}]{Ge:2016dlx}
\bibinfo{author}{\bibfnamefont{S.-F.} \bibnamefont{Ge}} \bibnamefont{and}
  \bibinfo{author}{\bibfnamefont{A.~{\relax Yu}.} \bibnamefont{Smirnov}},
  ``{Non-standard interactions and the CP phase measurements in neutrino
  oscillations at low energies},'' \bibinfo{journal}{JHEP}
  \textbf{\bibinfo{volume}{10}}, \bibinfo{pages}{138} (\bibinfo{year}{2016}),
  \eprint{1607.08513}.

\bibitem[{\citenamefont{Fukasawa
  et~al.}(2016{\natexlab{b}})\citenamefont{Fukasawa, Ghosh, and
  Yasuda}}]{Fukasawa:2016lew}
\bibinfo{author}{\bibfnamefont{S.}~\bibnamefont{Fukasawa}},
  \bibinfo{author}{\bibfnamefont{M.}~\bibnamefont{Ghosh}}, \bibnamefont{and}
  \bibinfo{author}{\bibfnamefont{O.}~\bibnamefont{Yasuda}}, ``{Sensitivity of
  the T2HKK experiment to the non-standard interaction},''
  (\bibinfo{year}{2016}{\natexlab{b}}), \eprint{1611.06141}.

\bibitem[{\citenamefont{Fukasawa and Yasuda}(2017)}]{Fukasawa:2016nwn}
\bibinfo{author}{\bibfnamefont{S.}~\bibnamefont{Fukasawa}} \bibnamefont{and}
  \bibinfo{author}{\bibfnamefont{O.}~\bibnamefont{Yasuda}}, ``{The possibility
  to observe the non-standard interaction by the Hyperkamiokande atmospheric
  neutrino experiment},'' \bibinfo{journal}{Nucl. Phys.}
  \textbf{\bibinfo{volume}{B914}}, \bibinfo{pages}{99} (\bibinfo{year}{2017}),
  \eprint{1608.05897}.

\bibitem[{\citenamefont{Liao et~al.}(2017)\citenamefont{Liao, Marfatia, and
  Whisnant}}]{Liao:2016orc}
\bibinfo{author}{\bibfnamefont{J.}~\bibnamefont{Liao}},
  \bibinfo{author}{\bibfnamefont{D.}~\bibnamefont{Marfatia}}, \bibnamefont{and}
  \bibinfo{author}{\bibfnamefont{K.}~\bibnamefont{Whisnant}}, ``{Nonstandard
  neutrino interactions at DUNE, T2HK and T2HKK},'' \bibinfo{journal}{JHEP}
  \textbf{\bibinfo{volume}{01}}, \bibinfo{pages}{071} (\bibinfo{year}{2017}),
  \eprint{1612.01443}.

\bibitem[{\citenamefont{Rout et~al.}(2017)\citenamefont{Rout, Masud, and
  Mehta}}]{Rout:2017udo}
\bibinfo{author}{\bibfnamefont{J.}~\bibnamefont{Rout}},
  \bibinfo{author}{\bibfnamefont{M.}~\bibnamefont{Masud}}, \bibnamefont{and}
  \bibinfo{author}{\bibfnamefont{P.}~\bibnamefont{Mehta}}, ``{Can we probe
  intrinsic CP/T violation and non-unitarity at long baseline accelerator
  experiments?},''  (\bibinfo{year}{2017}), \eprint{1702.02163}.

\bibitem[{\citenamefont{Ghosh and Yasuda}(2017)}]{Ghosh:2017ged}
\bibinfo{author}{\bibfnamefont{M.}~\bibnamefont{Ghosh}} \bibnamefont{and}
  \bibinfo{author}{\bibfnamefont{O.}~\bibnamefont{Yasuda}}, ``{Effect of
  systematics in T2HK, T2HKK and DUNE},''  (\bibinfo{year}{2017}),
  \eprint{1702.06482}.

\bibitem[{\citenamefont{Patrignani et~al.}(2016)}]{Olive:2016xmw}
\bibinfo{author}{\bibfnamefont{C.}~\bibnamefont{Patrignani}}
  \bibnamefont{et~al.} (\bibinfo{collaboration}{Particle Data Group}),
  ``{Review of Particle Physics},'' \bibinfo{journal}{Chin. Phys.}
  \textbf{\bibinfo{volume}{C40}}, \bibinfo{pages}{100001}
  (\bibinfo{year}{2016}).

\bibitem[{\citenamefont{Dziewonski and Anderson}(1981)}]{Dziewonski:1981xy}
\bibinfo{author}{\bibfnamefont{A.~M.} \bibnamefont{Dziewonski}}
  \bibnamefont{and} \bibinfo{author}{\bibfnamefont{D.~L.}
  \bibnamefont{Anderson}}, ``{Preliminary reference earth model},''
  \bibinfo{journal}{Phys. Earth Planet. Interiors}
  \textbf{\bibinfo{volume}{25}}, \bibinfo{pages}{297} (\bibinfo{year}{1981}).

\bibitem[{\citenamefont{Esteban et~al.}(2017)\citenamefont{Esteban,
  Gonzalez-Garcia, Maltoni, Martinez-Soler, and Schwetz}}]{Esteban:2016qun}
\bibinfo{author}{\bibfnamefont{I.}~\bibnamefont{Esteban}},
  \bibinfo{author}{\bibfnamefont{M.~C.} \bibnamefont{Gonzalez-Garcia}},
  \bibinfo{author}{\bibfnamefont{M.}~\bibnamefont{Maltoni}},
  \bibinfo{author}{\bibfnamefont{I.}~\bibnamefont{Martinez-Soler}},
  \bibnamefont{and} \bibinfo{author}{\bibfnamefont{T.}~\bibnamefont{Schwetz}},
  ``{Updated fit to three neutrino mixing: exploring the accelerator-reactor
  complementarity},'' \bibinfo{journal}{JHEP} \textbf{\bibinfo{volume}{01}},
  \bibinfo{pages}{087} (\bibinfo{year}{2017}), \eprint{1611.01514, NuFIT 3.0
  (2016), www.nu-fit.org}.

\bibitem[{\citenamefont{Adamson et~al.}(2016{\natexlab{a}})}]{Adamson:2016jku}
\bibinfo{author}{\bibfnamefont{P.}~\bibnamefont{Adamson}} \bibnamefont{et~al.}
  (\bibinfo{collaboration}{MINOS, Daya Bay}), ``{Limits on Active to Sterile
  Neutrino Oscillations from Disappearance Searches in the MINOS, Daya Bay, and
  Bugey-3 Experiments},'' \bibinfo{journal}{Phys. Rev. Lett.}
  \textbf{\bibinfo{volume}{117}}, \bibinfo{pages}{151801}
  (\bibinfo{year}{2016}{\natexlab{a}}), \bibinfo{note}{[Addendum: Phys. Rev.
  Lett.117,no.20,209901(2016)]}, \eprint{1607.01177}.

\bibitem[{\citenamefont{Ohlsson}(2013)}]{Ohlsson:2012kf}
\bibinfo{author}{\bibfnamefont{T.}~\bibnamefont{Ohlsson}}, ``Status of
  non-standard neutrino interactions,'' \bibinfo{journal}{Rept. Prog. Phys.}
  \textbf{\bibinfo{volume}{76}}, \bibinfo{pages}{044201}
  (\bibinfo{year}{2013}), \eprint{1209.2710}.

\bibitem[{\citenamefont{Kikuchi et~al.}(2009)\citenamefont{Kikuchi, Minakata,
  and Uchinami}}]{Kikuchi:2008vq}
\bibinfo{author}{\bibfnamefont{T.}~\bibnamefont{Kikuchi}},
  \bibinfo{author}{\bibfnamefont{H.}~\bibnamefont{Minakata}}, \bibnamefont{and}
  \bibinfo{author}{\bibfnamefont{S.}~\bibnamefont{Uchinami}}, ``{Perturbation
  Theory of Neutrino Oscillation with Nonstandard Neutrino Interactions},''
  \bibinfo{journal}{JHEP} \textbf{\bibinfo{volume}{03}}, \bibinfo{pages}{114}
  (\bibinfo{year}{2009}), \eprint{0809.3312}.

\bibitem[{\citenamefont{Biggio et~al.}(2009)\citenamefont{Biggio, Blennow, and
  Fernandez-Martinez}}]{Biggio:2009nt}
\bibinfo{author}{\bibfnamefont{C.}~\bibnamefont{Biggio}},
  \bibinfo{author}{\bibfnamefont{M.}~\bibnamefont{Blennow}}, \bibnamefont{and}
  \bibinfo{author}{\bibfnamefont{E.}~\bibnamefont{Fernandez-Martinez}},
  ``{General bounds on non-standard neutrino interactions},''
  \bibinfo{journal}{JHEP} \textbf{\bibinfo{volume}{08}}, \bibinfo{pages}{090}
  (\bibinfo{year}{2009}), \eprint{0907.0097}.

\bibitem[{\citenamefont{Gonzalez-Garcia and
  Maltoni}(2013)}]{Gonzalez-Garcia:2013usa}
\bibinfo{author}{\bibfnamefont{M.~C.} \bibnamefont{Gonzalez-Garcia}}
  \bibnamefont{and} \bibinfo{author}{\bibfnamefont{M.}~\bibnamefont{Maltoni}},
  ``{Determination of matter potential from global analysis of neutrino
  oscillation data},'' \bibinfo{journal}{JHEP} \textbf{\bibinfo{volume}{09}},
  \bibinfo{pages}{152} (\bibinfo{year}{2013}), \eprint{1307.3092}.

\bibitem[{\citenamefont{Formaggio and Zeller}(2012)}]{Formaggio:2013kya}
\bibinfo{author}{\bibfnamefont{J.~A.} \bibnamefont{Formaggio}}
  \bibnamefont{and} \bibinfo{author}{\bibfnamefont{G.~P.}
  \bibnamefont{Zeller}}, ``{From eV to EeV: Neutrino Cross Sections Across
  Energy Scales},'' \bibinfo{journal}{Rev. Mod. Phys.}
  \textbf{\bibinfo{volume}{84}}, \bibinfo{pages}{1307} (\bibinfo{year}{2012}),
  \eprint{1305.7513}.

\bibitem[{\citenamefont{Abe et~al.}(2015{\natexlab{b}})}]{Abe:2014gda}
\bibinfo{author}{\bibfnamefont{K.}~\bibnamefont{Abe}} \bibnamefont{et~al.}
  (\bibinfo{collaboration}{Super-Kamiokande}), ``{Limits on sterile neutrino
  mixing using atmospheric neutrinos in Super-Kamiokande},''
  \bibinfo{journal}{Phys. Rev.} \textbf{\bibinfo{volume}{D91}},
  \bibinfo{pages}{052019} (\bibinfo{year}{2015}{\natexlab{b}}),
  \eprint{1410.2008}.

\bibitem[{\citenamefont{Honda et~al.}(2015)\citenamefont{Honda, Sajjad~Athar,
  Kajita, Kasahara, and Midorikawa}}]{Honda:2015fha}
\bibinfo{author}{\bibfnamefont{M.}~\bibnamefont{Honda}},
  \bibinfo{author}{\bibfnamefont{M.}~\bibnamefont{Sajjad~Athar}},
  \bibinfo{author}{\bibfnamefont{T.}~\bibnamefont{Kajita}},
  \bibinfo{author}{\bibfnamefont{K.}~\bibnamefont{Kasahara}}, \bibnamefont{and}
  \bibinfo{author}{\bibfnamefont{S.}~\bibnamefont{Midorikawa}}, ``{Atmospheric
  neutrino flux calculation using the NRLMSISE-00 atmospheric model},''
  \bibinfo{journal}{Phys. Rev.} \textbf{\bibinfo{volume}{D92}},
  \bibinfo{pages}{023004} (\bibinfo{year}{2015}), \eprint{1502.03916}.

\bibitem[{\citenamefont{Foreman-Mackey
  et~al.}(2013)\citenamefont{Foreman-Mackey, Hogg, Lang, and
  Goodman}}]{ForemanMackey:2012ig}
\bibinfo{author}{\bibfnamefont{D.}~\bibnamefont{Foreman-Mackey}},
  \bibinfo{author}{\bibfnamefont{D.~W.} \bibnamefont{Hogg}},
  \bibinfo{author}{\bibfnamefont{D.}~\bibnamefont{Lang}}, \bibnamefont{and}
  \bibinfo{author}{\bibfnamefont{J.}~\bibnamefont{Goodman}}, ``{emcee: The MCMC
  Hammer},'' \bibinfo{journal}{Publ. Astron. Soc. Pac.}
  \textbf{\bibinfo{volume}{125}}, \bibinfo{pages}{306} (\bibinfo{year}{2013}),
  \eprint{1202.3665}.

\bibitem[{\citenamefont{Adamson et~al.}(2016{\natexlab{b}})}]{MINOS:2016viw}
\bibinfo{author}{\bibfnamefont{P.}~\bibnamefont{Adamson}} \bibnamefont{et~al.}
  (\bibinfo{collaboration}{MINOS}), ``{Search for Sterile Neutrinos Mixing with
  Muon Neutrinos in MINOS},'' \bibinfo{journal}{Phys. Rev. Lett.}
  \textbf{\bibinfo{volume}{117}}, \bibinfo{pages}{151803}
  (\bibinfo{year}{2016}{\natexlab{b}}), \eprint{1607.01176}.

\bibitem[{\citenamefont{Aartsen et~al.}(2016)}]{TheIceCube:2016oqi}
\bibinfo{author}{\bibfnamefont{M.~G.} \bibnamefont{Aartsen}}
  \bibnamefont{et~al.} (\bibinfo{collaboration}{IceCube}), ``{Searches for
  Sterile Neutrinos with the IceCube Detector},'' \bibinfo{journal}{Phys. Rev.
  Lett.} \textbf{\bibinfo{volume}{117}}, \bibinfo{pages}{071801}
  (\bibinfo{year}{2016}), \eprint{1605.01990}.

\bibitem[{\citenamefont{Kopp et~al.}(2013)\citenamefont{Kopp, Machado, Maltoni,
  and Schwetz}}]{Kopp:2013vaa}
\bibinfo{author}{\bibfnamefont{J.}~\bibnamefont{Kopp}},
  \bibinfo{author}{\bibfnamefont{P.~A.~N.} \bibnamefont{Machado}},
  \bibinfo{author}{\bibfnamefont{M.}~\bibnamefont{Maltoni}}, \bibnamefont{and}
  \bibinfo{author}{\bibfnamefont{T.}~\bibnamefont{Schwetz}}, ``{Sterile
  Neutrino Oscillations: The Global Picture},'' \bibinfo{journal}{JHEP}
  \textbf{\bibinfo{volume}{05}}, \bibinfo{pages}{050} (\bibinfo{year}{2013}),
  \eprint{1303.3011}.

\bibitem[{\citenamefont{Mitsuka et~al.}(2011)}]{Mitsuka:2011ty}
\bibinfo{author}{\bibfnamefont{G.}~\bibnamefont{Mitsuka}} \bibnamefont{et~al.}
  (\bibinfo{collaboration}{Super-Kamiokande}), ``{Study of Non-Standard
  Neutrino Interactions with Atmospheric Neutrino Data in Super-Kamiokande I
  and II},'' \bibinfo{journal}{Phys. Rev.} \textbf{\bibinfo{volume}{D84}},
  \bibinfo{pages}{113008} (\bibinfo{year}{2011}), \eprint{1109.1889}.

\bibitem[{\citenamefont{Day}(2016)}]{Day:2016shw}
\bibinfo{author}{\bibfnamefont{M.}~\bibnamefont{Day}}
  (\bibinfo{collaboration}{IceCube}), ``{Non-standard neutrino interactions in
  IceCube},'' \bibinfo{journal}{J. Phys. Conf. Ser.}
  \textbf{\bibinfo{volume}{718}}, \bibinfo{pages}{062011}
  (\bibinfo{year}{2016}).

\end{thebibliography}

\end{document}